\documentclass[journal,12pt, onecolumn]{IEEEtran}
\usepackage{etex}
\usepackage{array,multirow}

\usepackage{floatrow}
\usepackage{adjustbox}
\usepackage{amsthm}
\usepackage[utf8]{inputenc}
\usepackage[T1]{fontenc}
\usepackage{ifthen}
\usepackage{cite}
\usepackage{graphicx}
\usepackage[cmex10]{amsmath}
\usepackage{amsfonts,amssymb}
\usepackage{color}
\usepackage{ifthen}
\usepackage{accents}
\usepackage[ruled,vlined]{algorithm2e}
\usepackage{listings}
\usepackage{footnote}
\usepackage{array}
\usepackage{verbatim, animate}
\usepackage{comment}
\usepackage[mathscr]{euscript}
\usepackage{amsfonts,amssymb,latexsym, amsxtra,enumerate}
\usepackage[all]{xy}
\usepackage{psfrag}
\usepackage{xcolor}
\usepackage{lineno}
\usepackage[font=footnotesize]{caption}
\usepackage{multirow}

\usepackage{tikz}
\usetikzlibrary{positioning}
\newfloatcommand{capbtabbox}{table}[][\FBwidth]

\newcommand{\bld}[1]{\mathbf{#1}}

\newlength{\dhatheight}

\newcommand{\bm}[1]{\begin{bmatrix}#1\end{bmatrix}}

\newcommand{\figref}[1]{Fig.~\ref{#1}}
\newcommand{\sect}[1]{Section~\ref{#1}}

\newcommand{\mE}{\mathbb{E}}
\newcommand{\mR}{\mathbb{R}}

\newcommand{\sq}[1]{\left[#1\right]}
\newcommand{\cq}[1]{\left(#1\right)}
\newcommand{\bq}[1]{\left\{#1\right\}}

\newcommand{\gtt}{\gamma_{t+1}}

\newcommand{\pushright}[1]{\ifmeasuring@ #1 \else\omit\hfill$\displaystyle#1$\fi\ignorespaces}
\newcommand{\pushleft}[1]{\ifmeasuring@ #1 \else\omit$\displaystyle#1$\hfill\fi\ignorespaces}

\newcommand{\phienc}{\phi_{t}^{\mathrm{enc}}}
\newcommand{\psidec}{\psi^{\mathrm{dec}}}
\newcommand{\at}{\tilde{a}}
\newcommand{\tphi}{\tilde{\phi}}
\newcommand{\sut}{\sigma_{u,t}}
\newcommand{\sutt}{\sigma_{u,t+1}}

\newtheorem{lemma}{Lemma}

\newtheorem{theorem}{Theorem}
\newtheorem{remark}{Remark}

\newtheorem{corollary}{Corollary}
\newtheorem{definition}{Definition}

\newcommand{\Ncal}{\mathcal{N}}

\DeclareMathOperator\E{\textsf{E}}

\def\gt{\gamma_t}

\makeatletter
\newcommand{\printfnsymbol}[1]{%
  \textsuperscript{\@fnsymbol{#1}}%
}
\makeatother

\newcommand{\yt}{{\tilde{y}}}
\newcommand{\zt}{{\tilde{z}}}
\newcommand{\nt}{{\tilde{n}}}
\usepackage{xspace}
\usepackage[acronym,nogroupskip,nonumberlist,nopostdot]{glossaries}

\newacronym{mdp}{MDP}{Markov decision process}
\newacronym{ne}{NE}{Nash equillibrium}
\newacronym{mfe}{MFE}{mean-field equillibrium}
\newacronym{mpe}{MPE}{Markov perfect equillibrium}
\newacronym{mfg}{MFG}{mean-field game}
\newacronym{rl}{RL}{reinforcement learning}
\newacronym{marl}{MARL}{multi-agent reinforcement learning}
\newacronym{iot}{IoT}{Internet of Things}
\newacronym{ssg}{SSG}{Stackelberg Security Game}
\newacronym{pse}{PSE}{Perfect Stackelberg Equilibrium}
\newacronym{mpse}{MPSE}{Markov \gls{pse}}
\newacronym{spbe}{SPBE}{Sub-game Perfect Bayesian Equilibrium}
\newacronym{pf}{PF}{particle filter}
\newacronym{pomdp}{POMDP}{partially obseravable \gls{mdp}}
\newacronym{nn}{NN}{neural network}
\newacronym{td}{TD}{Temporal Difference}
\newacronym{awgn}{AWGN}{Additive White Gaussian noise}
\newacronym{tim}{TIM}{topological interference management}
\newacronym{gdof}{GDoF}{generalized degrees of freedom}
\newacronym{snr}{SNR}{Signal to Noise Ratio}
\newacronym{sinr}{SINR}{signal to noise and interference ratio}
\newacronym{inr}{INR}{interference to noise ratio}
\newacronym{pam}{PAM}{Pulse Amplitude Modulation}
\newacronym{mse}{MSE}{Mean Squared Error}
\newacronym{ber}{BER}{Bit Error Rate}
\newacronym{tdm}{TDM}{time domain multiplexing}
\newacronym{gan}{GAN}{Generative Adversarial Network}
\newacronym{sir}{SIR}{signal to interference ratio}
\newacronym{mmse}{MMSE}{Minimum Mean-squared error}
\newacronym{sk}{SK}{Schalkwijk-Kailath}
\newacronym{bsc}{BSC}{binary symmetric channel}
\newacronym{ldpc}{LDPC}{low density parity codes}
\newacronym{ml}{ML}{maximum likelihood}
\newacronym{dp}{DP}{Dynamic Programming}
\newacronym{rnn}{RNN}{Recurrent Neural Network}
\newacronym{cdf}{CDF}{Cummulative Distribution Function}
\newacronym{cl}{CL}{Chance and Love}

\newcommand{\cl}{\gls{cl}\xspace}

\newcommand{\dpp}{\gls{dp}\xspace}

\newcommand{\bsc}{\gls{bsc}\xspace}
\newcommand{\sk}{\gls{sk}\xspace}
\newcommand{\mdp}{\gls{mdp}\xspace}

\newcommand{\awgn}{\gls{awgn}\xspace}

\newcommand{\snr}{\gls{snr}\xspace}

\newcommand{\pam}{\gls{pam}\xspace}
\newcommand{\mse}{\gls{mse}\xspace}

\newcommand{\mmse}{\gls{mmse}\xspace}
\newcommand{\cdf}{\gls{cdf}\xspace}
\usepackage{algorithmic}
\usepackage[ruled,vlined]{algorithm2e}

\allowdisplaybreaks

\IEEEoverridecommandlockouts
\makeatletter

\title{Linear Coding for AWGN channels with Noisy Output Feedback via Dynamic Programming}

\author{Rajesh Mishra\textsuperscript{1,*}, Deepanshu Vasal\textsuperscript{2,*} and Hyeji Kim\textsuperscript{1}
\thanks{$^1$ The University of Texas at Austin}
\thanks{$^2$ Northwestern University}
 \thanks{$^*$ indicates authors have equal contribution}
 \thanks{This paper was presented in part at Proc. IEEE Int. Symp. Inf. Theory, 2021
 ~\cite{MiVaKi21}.}
}

\begin{document}

\maketitle

\begin{abstract}
  The optimal coding scheme for communicating a Gaussian message over an \awgn channel with \emph{\awgn output feedback}, with a limited number of transmissions, is unknown. Even if we restrict the scope of the coding scheme to linear schemes, still, deriving the optimal coding scheme is a challenging task. The state-of-the-art linear scheme for channels with noisy feedback is by Chance and Love, where the coefficients of the linear scheme are numerically optimized based on unique observations~\cite{Chance2011a}. In this paper, we introduce a new class of linear coding schemes, which we call sequential linear schemes, for this channel by introducing a linear state process at the transmitter. We then derive the optimal scheme within this class in a closed-form by formulating a novel \dpp. We empirically show that our scheme outperforms the state-of-the-art linear scheme in~\cite{Chance2011a} for noisy feedback and coincides with the Shalkwijk-Kailath scheme for noiseless feedback. This problem is an instance of decentralized control \emph{without any common information} and to the best of our knowledge the first such scenario where we can derive analytical solutions using a \dpp. Finally, we show that in communicating message {\em bits} instead of a Gaussian message, a learning-based approach further improves the reliability of sequential linear schemes. 
\end{abstract}

\section{Introduction}

The study of channels with {\em output feedback} was initiated by Shannon~\cite{Shannon1956}, where he shows that the feedback of the output
does not increase the capacity of point to point \awgn channels and any discrete memoryless channels. Despite such negative results,
feedback is shown to improve the reliability in the {\em finite blocklength} regime~\cite{Horstein1963, Schalkwijk1966a, Schalkwijk1966b,shayevitz2010optimal}. For channels with {\em noiseless} output feedback, Horstein studied \bsc channel and presented a scheme that achieves capacity~\cite{Horstein1963}. Schalkwijk and Kailath in~\cite{Schalkwijk1966a, Schalkwijk1966b} studied \awgn channel and proposed an optimal linear coding scheme that achieves capacity and a doubly exponential error exponent~\cite{Schalkwijk1966a}. Both Horstein and \sk schemes were later generalized by Shayevitz and Feder by proposing a posterior matching scheme~\cite{shayevitz2010optimal} for an arbitrary discrete memoryless channel where the encoder transmits the generalized inverse of the capacity-achieving input \cdf at every transmission.

For channels with {\em noisy} output feedback, on the other hand, far less is known. The celebrated \sk scheme does not readily generalize to noisy feedback channels~\cite{shayevitz2010optimal}. Chance and Love proposed a linear scheme that significantly outperforms the \sk scheme for \awgn channels with \awgn noisy feedback~\cite{Chance2011a}. They also introduced a concatenated coding scheme where their linear scheme is used as an inner code and the forward error-correcting code is used as an outer code. Nevertheless, whether the \cl scheme is optimal has remained unknown over the last decade. We make progress on this long-standing open problem; we derive a linear coding scheme that outperforms the CL scheme for channels with noisy output feedback.

 While we focus on \awgn channels with \awgn noisy {output} feedback, we note that there are various other models for channels with feedback. In \cite{martins2008coding}, Martins and Weissman consider a channel where the feedback is affected by quantization noise or an additive bounded noise and provide a scheme that performs close to capacity. In \cite{burnashev2008bsc}, Burnashev and Yamamoto consider a \bsc as both forward as well as the feedback channel and show that for some positive crossover probability of the feedback channel, the achievable error exponent outperforms that of the no-feedback case. This result was further refined in~\cite{Hirosuke2014} and extended by Xiang and Kim to show that the error exponent of channels with feedback is strictly higher compared to channels with no feedback if the noise variance of the feedback channel is sufficiently small~\cite{Young2010}. The error exponents for channels with noisy feedback for different power constraint assumptions on the feedback channel are derived by Kim and Weissman~\cite{Kim2011}. For channels with {\em active} feedback, where the receiver is allowed to actively {\em encode} its outputs, Ben-Yishai and Shayevitz introduce the Modulo-SK scheme which is shown to improve the reliability upon the \sk scheme by posing the problem as a joint source-channel coding with side information and solving it using an interactive modulo-lattice solution~\cite{Ben-Yishai2015}.

In general, communication with \emph{noiseless} feedback can be posed as a multi-user decentralized stochastic control problem~\cite{Coleman09} with common information at the transmitter and the receiver. In general, there is a conceptual framework to solve such problems within the framework of dynamic programming~\cite{Nayyar13}. However, when there is no common information among the decision-makers, there is no such framework available, mainly because in such problems each player needs to put a belief on other players' private information, and a belief on their beliefs ad infinitum. This is referred to as the infinite regress of higher-order beliefs in both control and economics literature. Point-to-point channel with noisy feedback is one such decentralized control problem without any common information and has thus lacked any mathematical framework to study this problem. However, recently, in ~\cite{Vasal2020}, Vasal presented a sequential decomposition methodology to decompose a general discrete memoryless point-to-point channel with noisy feedback by providing a notion of the state of this channel based on an auxiliary controller at the transmitter.

In this paper, we consider an \awgn channel with \awgn noisy output feedback and extend the notion of the state mentioned above to propose a \dpp algorithm that solves for the coefficients of a linear encoding scheme in closed-form. More specifically we introduce a class of linear schemes, called {\em sequential linear} schemes, and derive the optimal solution within this class for \awgn channels with noisy and noiseless feedback. For channels with noiseless feedback, we recover the \sk scheme as the optimal scheme in our framework as a special case. Note that \sk scheme was inspired by Robbin's scheme and was used in \awgn channel based on human intuition and there was no analytical framework to \emph{derive} the celebrated SK scheme. For \awgn channels with noisy feedback, surprisingly, we show that our sequential linear coding scheme strictly outperforms the CL scheme. Our main contributions are as follows:
%%\footnote}

\begin{itemize}
  \item We introduce a family of sequential linear schemes that are naturally equipped with a recursive optimal decoder, i.e., Kalman linear \mmse filter, for \awgn channels with noisy output feedback (\sect{sec: linear schemes}). We derive the closed-form solution for the optimal sequential linear code via dynamic programming under a constant peak power constraint. To do so, we introduce a novel \mdp framework that uses variances in the estimation at the transmitter and receiver as states (\sect{sec: dp}).

  \item We characterize the \mmse that the optimal schemes achieve as a function of the number of transmissions in a closed-form solution. We observe that the \mmse approximately drops exponentially for noiseless feedback while the drop in variance is polynomial for noisy feedback settings (\sect{sec: schemes}).

  \item We extend the results to a total power constraint. We provide a dynamic program to compute a linear scheme with the optimal power allocations for each of the transmissions for both noiseless and noisy feedback settings. We show that our scheme outperforms the state-of-the-art scheme for noisy feedback channels (\sect{sec: total power}).

  \item We evaluate the performance of our proposed sequential linear coding schemes for communicating message bits over \awgn channels with noisy output feedback. We conclude that for codes with finite constellation, the sequential linear coding scheme derived using the \dpp is sub-optimal. We develop a learning-based approach that improves the performance of sequential linear schemes for such cases (\sect{sec: PAM}).

\end{itemize}

\section{Problem setup and prior work}
\label{sec: Gauss Model}

\begin{figure}[!htb]
  \centering
  \includegraphics[width=.6\textwidth]{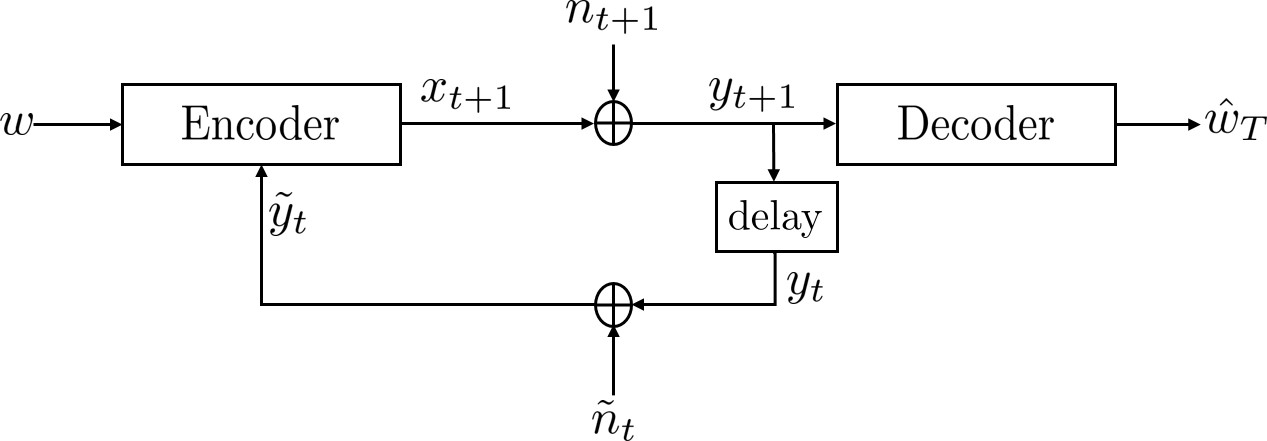}\ \
  \caption{\awgn channel with noisy output feedback}
  \label{fig:transmit_model}
\end{figure}

We consider \awgn channels with noisy output feedback, where the transmitter aims to communicate a message $W\sim \mathcal{N}\cq{0,\sigma_w^2}$ to the receiver over $T$ rounds of communications. As depicted in~\figref{fig:transmit_model}, the forward communication channel is modeled as an \awgn channel,
\begin{align*}
  y_t = x_t + n_t,
\end{align*}
where $x_t$ denotes the transmitted symbol, and $N_t \sim \Ncal(0,\sigma_f^2)$ denotes the additive Gaussian noise for $t=0,1,\cdots,T$. Each forward transmission is followed by an output feedback, where the transmitter receives
\begin{align*}
  \yt_t = y_t + \tilde{n}_t,
\end{align*}
where $\tilde{N}_t \sim \Ncal(0,\sigma_b^2)$ denotes the additive Gaussian noise in the feedback channel.

The encoding process is inherently causal. For every transmission except the initial transmission, the output feedback $\yt^t=\bq{\yt_0\cdots \yt_t}$ is used to frame the next transmission symbol as
\begin{align*}
  x_{t} = \phienc\cq{w,\yt^{t-1}}
\end{align*}
for $t \in [1,T]$, where the transmission power is constrained either in the peak power or total power. For the peak power constraint, we let $\E[X_t^2] \le P \ \forall\ t \in [0,T]$. For the total power constraint, we let $\E[\sum_{t=0}^{T} X_t^2] \le (T+1)P$. The decoding process, on the other hand, does not need to be causal; after $T+1$ rounds of transmissions, the decoder generates an estimated message $\hat{w}_T$ based on the entire received sequence $y^T = (y_0,\cdots,y_T)$ as
\begin{align*}
  \hat{w}_T = \psidec(y^T).
\end{align*}
Hence, designing a coding scheme for channels that utilizes output feedback involves jointly designing multiple encoding functions $\{\phienc(\cdot)\}$ for $t \in [1,T]$ and $\psidec(\cdot)$ that minimize the \mmse defined as $\E\sq{\cq{W-\hat{W}_T}^2}$.

Designing a coding scheme for channels with feedback is challenging due to ($a$) the necessity to optimize the encoder and decoder jointly and ($b$) the high dimensional encoding space. Thus, literature has focused on linear coding schemes, for which the transmitted symbol is a linear function of the message and the output feedback, formally defined in Definition~\ref{def: linear} and in~\cite{Chance2011a}. Linear schemes allow theoretical analysis and efficient implementation; they are naturally equipped with an optimal decoder, i.e., the linear \mmse estimator, assuming that the message $w$ is Gaussian, which eases the design of linear schemes.
\begin{definition}[Linear schemes]\label{def: linear}
  \textnormal{A sequence of schemes is called linear if the encoding function can be expressed as}
  \begin{align*}
    \phienc\cq{w, \yt^{t-1}} = a_t w + \sum_{j = 0}^{t-1} b_{t,j} \yt_j
  \end{align*}
  \textnormal{for some $a_t, b_{t,0}, \cdots, b_{t,t-1} \in \mR$ for every $t \in [0,T]$.}
\end{definition}
Nevertheless, there has been limited success in deriving the optimal linear schemes for \awgn channels with output feedback. For \awgn channels with noiseless feedback, the celebrated \sk scheme is shown to be optimal among all linear schemes~\cite{liu2006fundamental}. For \awgn channels with noisy output feedback, on the other hand, the optimal linear coding scheme is still unknown~\cite{Chance2011a}.
In the following, we review the known results on the linear coding for \awgn channels with output feedback. We begin by reviewing the Schalkwijk and Kailath~\cite{Schalkwijk1966}, which is the optimal linear coding scheme for channels with noiseless output feedback~\cite{Schalkwijk1966} (\sect{sec: sk}). We then review the best known linear coding scheme for channels with {\em noisy} feedback by Chance and Love~\cite{Chance2011a} (\sect{sec: cl}).
\subsection{Prior work on noiseless feedback: Optimal linear scheme by Schalkwijk and Kailath}
\label{sec: sk}

\sk scheme introduced by Schalkwijk and Kailath in~\cite{Schalkwijk1966} is a linear coding scheme for channels with {\em noiseless} feedback (i.e., $\tilde{y}_t = y_t$) that is shown to be optimal~\cite{butman1967optimum}. In the \sk scheme, as illustrated in Algorithm~\ref{alg: SK Scheme}, the encoder transmits its raw message $w$ in the first transmission. Afterwards, the encoder computes the error between $w$ and the receiver's \mmse estimate $\hat{w}_t:=\E[W|y^t]$ and transmits the error (with a scaling to satisfy the power constraint), i.e., $x_{t+1} \propto (w - \hat{w}_t)$.

\begin{algorithm}[!htb]
  \SetAlgoLined
  \DontPrintSemicolon
  \KwIn{$w$}
  \KwOut{Final Estimate $\hat{w}_T = \E[W|y_0,\cdots,y_T]$}
  \underline{Encoder:} \\
  $\indent$ Initialize $u_0 = w$ and $\gamma_0 = \sqrt{P/\E[W^2]}$\;
  $\indent$ $x_0 = \gamma_0 u_0$ \;
  \underline{Decoder:} \\
  $\indent$ $y_0 = x_0 + n_0$ \;
  \For{$t = 0, \cdots, T$}{
    \underline{Encoder;}\\
    $\indent$ %State update: 
    $u_{t+1} = u_t - \E[U_t|y_t] $ \;

    $\indent$ %Symbol generation:
    $x_{t+1} = \gamma_{t+1} u_{t+1}$, where $\gamma_{t+1} = \sqrt{P/\E[U_{t+1}^2]}$\;

    \underline{Decoder:}\\
    $\indent$
    $y_{t+1} = x_{t+1} + n_{t+1}$\;

  }
  \underline{Decoder:} $\hat{w}_T = \sum_{t=0}^T \E[U_t|y_t] $.
  \caption{Schalkwijk-Kailath scheme}
  \label{alg: SK Scheme}
\end{algorithm}

Extending the \sk scheme to channels with noisy output feedback, however, is not straightforward since the encoder is not aware of the received values $y^t$ and thus cannot compute the receiver's \mmse estimate $\hat{w}_t = \E[W|y^t]$.

\subsection{Prior work on noisy feedback: linear coding scheme by Chance and Love}
\label{sec: cl}
For \awgn channels with output feedback, Chance and Love in~\cite{Chance2011a} introduce a {\em concatenated} coding scheme, which consists of a linear code as the inner code concatenated with an error-correcting code as the outer code. In the following, we describe the methodology proposed in~\cite{Chance2011a} to optimize the linear encoding scheme for noisy feedback. The authors begin with the most general assumption of the linear feedback scheme given as
\begin{align*}
  \mathbf{x} & = \mathbf{F\cq{n+\tilde{n}}}+\mathbf{g}w,
\end{align*}
where $w$ is the message, $\bld{g}\in\mR^T$ is a unit-norm vector, $\bld{F}$ is a $T\times T$ lower triangular encoding matrix, $\bld{\tilde{n}+n}$ is the combined $T\times 1$ noise vector, and $\bld{x}$ is the final $T\times 1$ vector to be transmitted. At the receiver, a combination vector $\bld{q}\in\mR^T$ is used to extract the message $\hat{w}_T$ out of all the received symbols $\bld{y}$ expressed as
\begin{align*}
  \hat{w}_T & =\mathbf{q}^T\mathbf{y}.
\end{align*}

In order to find the optimal set of parameters $\bld{F}$, $\bld{g}$, and $\bld{q}$, one could represent the received \snr in terms of these parameters and then find $\bld{F}$, $\bld{g}$, and $\bld{q}$ that maximizes the received \snr. This optimization, however, is intractable. To mitigate this challenge, Chance and Love first show that for a given $\bld{F}$ and $\bld{g}$, the optimal vector $\bld{q}$ can be obtained in a closed form, which leaves two variables $\bld{F}$ and $\bld{g}$ to optimize. However, simultaneous optimization of the remaining two variables is still intractable; they propose an iterative approach wherein one variable is optimized, keeping the other fixed and vice versa.

In~\cite{Chance2011a}, they provide a concatenated code with an inner and an outer code. The inner code is the linear coding scheme using the optimized variables $\bld{F}$, $\bld{g}$ and $\bld{q}$. Nevertheless, the CL scheme we refer to in this paper does not include the concatenation (as the concatenation can be applied to any inner coding scheme). The CL scheme is shown to significantly outperform the \sk scheme for \awgn channels with noisy output feedback and coincides with the \sk scheme for \awgn channels with noiseless output feedback, which shows that a tailored coding scheme for channels with noisy output feedback does provide an additional reliability gain. They showcase an improvement of nearly $10$ dB of received \snr compared to the \sk scheme for certain settings~\cite{Chance2011a}.

Two very interesting questions, following Chance and Love's work, are ($a$) whether the CL scheme is optimal within the class of linear schemes and ($b$) whether one could derive a closed-form linear code. The CL scheme includes an iterative update of the matrices $\bld{F}$ and $\bld{q}$ and thus does not provide a closed-form expression. In the rest of the paper, we provide an answer to both questions. We derive a linear coding scheme with a closed-form expression that strictly generalizes the CL scheme, i.e., the proposed scheme is equal to or strictly more reliable than the CL scheme. As we elaborate on in the following section, our scheme is inspired by the CL scheme, which has a sequential structure.

\section{Sequential linear schemes}
\label{sec: linear schemes}

\begin{figure}[!htb]
  \centering
  \includegraphics[width=.5\textwidth]{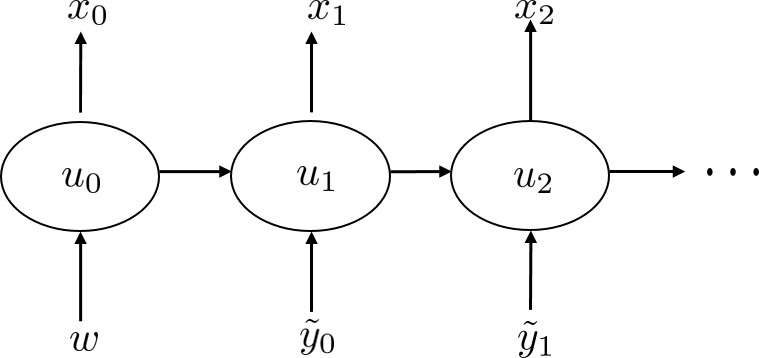}
  \caption{Sequential linear encoder for channels with noisy output feedback: The encoder state is updated as $u_{t+1} = b_t u_t + c_t \yt_t$  as in~\eqref{Eqn: Ut main update equation} and a scaled version of $u_t$ is sent as $x_t = \gamma_t u_t$}
  \label{fig: sequential linear code with ut state diagram}
\end{figure}

In this section, we introduce a new family of linear schemes, namely sequential linear schemes defined in Definition~\ref{Def: Seq lin}, and depicted in~\figref{fig: sequential linear code with ut state diagram}. We prove that the family of these sequential schemes, although a strict subset of linear schemes, strictly generalizes the \sk and CL schemes (Remarks~\ref{Remark: SK scheme as sequentially linear} and~\ref{Remark: chance and love as sequentially linear}). We then show that for the class of sequential linear schemes, a recursive Kalman filter can be used as an optimal estimator at the receiver (Section~\ref{sec: Estimation}), which is an essential precursor toward deriving the optimal sequential linear schemes.

\begin{definition}[Sequential linear schemes]\label{Def: Seq lin}
  \textnormal{
    A sequence of schemes is called {\em sequentially linear} if the encoder maintains a state $u_t \in \mR$ which is updated based on the most recent feedback $\yt_{t}$ in a linear manner as
    \begin{align}
      u_{t+1} = u_t + c_t \yt_t \label{Eqn: Ut main update equation}
    \end{align}
    for some $c_t \in \mathbb{R}$ and transmits a scaled version of $u_{t+1}$  as
    \begin{align}
      \label{Eqn: gamma definition}
      x_{t+1} = \gamma_{t+1} u_{t+1},
    \end{align}
    to satisfy the power constraint $\E[X_{t}^2] \le P$, for $t \in [0,T]$ under the peak power constraint or to satisfy the power constraint $\sum_{t=0}^{t=T}  \E[X_{t}^2] \le (T+1)P$ under the total power constraint.
  }
\end{definition}
For the convenience of notation, we let $\phi_t\cq{\cdot}$ denote the sequential linear encoding operation at time $t$ parameterized by $(\gamma_{t+1},c_t)$, i.e.,
\begin{align}
  x_{t+1} & = \phi_t\cq{u_t, \yt_t}
  =\gtt(u_t+c_t\yt_t).\label{Eqn: Encoder function}
\end{align}
\begin{remark} [Not all linear schemes are sequential]
  \textnormal
  {
    As we restrict the state $u_t$ to be a scalar, sequential linear schemes are a {\em strict} subset of linear schemes. In other words, not all linear schemes are sequential. On the other hand, if we allow $u_t$ to be a vector of length $T$, i.e., $u_t \in \mathbb{R}^T$, where $T$ denotes the length of total transmissions, then sequential linear schemes include the entire class of linear schemes.
  }
\end{remark}
While the restriction on the dimension of $u_t$ makes the sequential linear schemes a strict subset of linear schemes, in the following, we show that both the \sk scheme and the CL scheme belong to the family of sequential linear schemes.

\begin{remark}[SK scheme is sequentially linear]
  \label{Remark: SK scheme as sequentially linear}
  \textnormal
  {
    \sk scheme falls into the family of sequential linear schemes. Let $u_t$ denote the estimation error, i.e., $u_t = w - \hat{w}_t$. Then it follows that $u_{t+1} = u_t + c_t \tilde{y}_t$, where $c_t = -\E[U_t \tilde{Y}_t]/\E[\tilde{Y}_t^2]$ and $x_{t+1} = \gamma_{t+1} u_{t+1}$, where $\gamma_{t+1}$ is chosen to satisfy the power constraint $\E[X_{t+1}^2] \le P$ for every $t \in [0,T]$. (See Appendix~\ref{App: SK scheme as sequentially linear} for detailed proof.)
  }
\end{remark}

\begin{remark}[CL scheme is sequentially linear]
  \label{Remark: chance and love as sequentially linear}
  \textnormal
  {
    CL scheme starts with the general form of linear schemes, but the conjectured optimal linear schemes fall into the family of sequential linear encoding. (See Appendix~\ref{App: chance and love as sequentially linear} for detailed proof.)
  }
\end{remark}

\subsection{Linear \mmse Estimation}
\label{sec: Estimation}

In this section, we show that ($a$) sequential linear schemes are naturally equipped with an efficient and recursive \mmse estimator, namely, the Kalman filter, and ($b$) the \mmse can be represented as a function of the parameters of the encoding scheme. In other words, for any choice of $\{c_t, \gamma_t\}_{t=1}^T$ in the sequential linear encoding scheme, we can represent the \mmse as a function of $\{c_t, \gamma_t\}_{t=1}^T$. Kalman filter provides a recursive \mmse estimation of a state variable when the state variable follows a state-space equation, and the observation variable can be written as a sum of the observation variable and the noise.

In order to derive the recursive Kalman estimator for sequential linear schemes, we begin with the observation that the encoder's state $u_t$ satisfies the linear state space equation in~\eqref{Eqn: Ut main update equation} and $y_t$ is a scaled version of $u_t$ plus the Gaussian noise. Under these conditions, the Kalman filter allows one to recursively update the estimation of the {\em state} $u_t$ given a series of observations $y^t$. Nevertheless, this alone is insufficient to estimate the {\em message} $w$ given $y^t$. In order to derive a recursive estimation of $w$ given $y^t$, we let $p_t = \bm{w & u_t}^T$ denote the pair of the message $w$ and the encoder state $u_t$ as shown below.

\begin{lemma}[Kalman Filter as Decoder]
  \label{lemma:linearkalman}
  \textnormal
  {
    Let us define the Kalman state variable $p_t=\bm{w &u_t}^T$, where $w$ is the message and $u_t$ is the encoder state defined in the sequential linear encoding scheme at the transmitter from~\figref{fig: sequential linear code with ut state diagram}. Then, $p_t$ and $y_t$ satisfy the following
    \begin{align}
      p_{t+1} & = A_t p_t + G_t z_t,\;\;\;\; \label{Eqn: Kalman state equation}
      y_t = C_tp_t + n_t,
    \end{align}
    where the matrices $A_t$, $G_t$, $C_t$ are deduced from the equations representing the forward and the feedback transmission as
    \begin{align*}
      A_t = \bm{1 & 0 \\ 0 & 1 + \gamma_t c_t}, \qquad G_t =\bm{0 & 0 \\ c_t & c_t}, \qquad C_t =\bm{0 & \gt},
    \end{align*}
    and the vectors $z_t=\bm{n_t & \nt_t}$ and forward Gaussian noise $n_t$ represent the noises in the state update and the observations, respectively.
  }
  \begin{figure}[!htb]
    \centering
    \includegraphics[width=.5\textwidth]{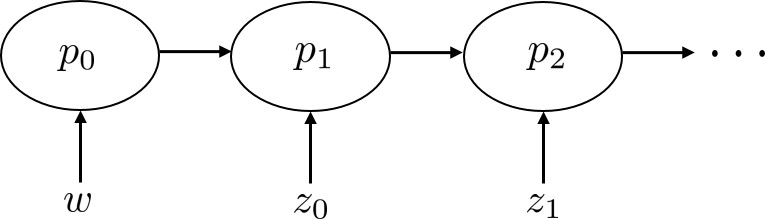}
    \caption{Sequential linear update of the state $p_t$ where $w$ is the message, $z_t = [n_t \nt_t]$ with $n_t$ and $\nt_t$ being the forward and the feedback noise respectively}
    \label{fig: sequential linear update of pt}
  \end{figure}
\end{lemma}

\begin{proof}
  Lemma~\ref{lemma:linearkalman} follows immediately from the definition of the sequential linear schemes in Definition~\ref{Def: Seq lin}.
\end{proof}

Based on Lemma~\ref{lemma:linearkalman}, we establish the following theorem.

\begin{theorem}[\mmse estimation via Kalman filter]
  \label{Thm: Kalman Filter Derivation}
  \textnormal
  {
    Let $p_t = \bm{w & u_t}^T$ denote the pair of the message and the $t^{th}$ encoder state. Let $\hat{p}_t := \E[P_t|y^t]$ denote its estimate and $\Sigma_t := \E[(P_t - \hat{P}_t)(P_t - \hat{P}_t)^T]$ denote the covariance of the corresponding error, given the observation vector $y^t = (y_0, \cdots, y_t)$. Then the estimate $\hat{p}_t$ and its error covariance matrix $\Sigma_t$ can be recursively computed via Kalman filter as
    \begin{align}
      \hat{p}_{t+1} = g_1(\hat{p}_t, \gt, \gtt, c_t, y_t), \qquad
      \Sigma_{t+1} = g_2(\Sigma_t, \gt, \gtt, c_t),\label{Eqn: pthat and Sigmat}
    \end{align}
    where $g_1(\cdot)$ and $g_2(\cdot)$ are defined in~\eqref{Eqn: estimation update} and~\eqref{Eqn: error covariance update}, respectively, and $\gt$ and $c_t$ denote the encoder parameters defined in Definition~\ref{Def: Seq lin}.
  }
\end{theorem}

\vspace{.5em}

\begin{proof}

  As the state variable $p_t$ and the observation variable $y_t$ follow the canonical form of equations in~\eqref{Eqn: Kalman state equation} as shown in Lemma~\ref{lemma:linearkalman}, the Kalman filter provides a recursive form of the \mmse estimate and the corresponding \mmse error covariance matrix.

  Let $\hat{p}_t := \E[P_t|y^t]$ denote the \mmse estimate of $p_t$ given $y^t$ and $\Sigma_t$ denote the corresponding error covariance matrix defined as
  \begin{align}
    \Sigma_{t} & := \mE\sq{\cq{P_{t}-\hat{P}_t}\cq{P_{t}-\hat{P}_t}^T} =\bm{\epsilon_{w,t}^2 & \epsilon_{uw,t}^2 \\ \epsilon_{uw,t}^2 &\epsilon_{u,t}^2 },\label{Eqn: Sigmat definition}
  \end{align}
  where $\epsilon_{w,t}^2 := \mE\sq{\cq{W-\hat{W}_t}^2}$ denotes the \mse in estimating the message, $\epsilon_{u,t}^2  := \mE\sq{\cq{U_t-\hat{U}_t}^2}$ denotes the \mmse in estimating the encoder's most recent state, and $\epsilon_{uw,t}^2 := \mE\sq{\cq{W - \hat{W}_t}\cq{U_t-\hat{U}_t}}$ denotes the covariance of the two errors.
  Using this canonical form and the standard definitions for Kalman filter solutions we can express the \mmse estimate of $\hat{p}_t$ in a recursive form as
  \begin{align}
    \hat{p}_{t+1} = A_t\hat{p}_t + L_{t+1}\cq{y_{t+1} - C_{t+1}A_t\hat{p}_t}, \label{Eqn: estimation update}
  \end{align}
  where
  \begin{align*}
    L_{t+1} = \frac{\cq{A_{t}\Sigma_{t}A_{t}^T+ G_tQG_t^T}C_{t+1}^T}{C_{t+1}\cq{A_{t}\Sigma_{t}A_{t}^T+ G_tQG_t^T}C_{t+1}^T + \sigma_f^2} \qquad \text{and} \qquad Q=\E[Z_t Z_t^T] = \bm{\sigma_f^2 & 0 \\ 0 & \sigma_b^2}.
  \end{align*}

  Similarly, $\Sigma_t$ follows a recursive relation given as
  \begin{align}
    \Sigma_{t+1} & = \cq{I-L_{t+1}C_{t+1}}\cq{A_{t}\Sigma_{t}A_{t}^T+ G_tQG_t^T}, \label{Eqn: error covariance update}
  \end{align}
  where $\hat{p}_{-1}$ and $\Sigma_{-1}$ are initialized as $\hat{p}_{-1} = \mathbf{0}$ and $\Sigma_{-1} = \bm{\sigma_w^2 & \sigma_w^2 \\ \sigma_w^2 & \sigma_w^2}.$
\end{proof}
\vspace{1em}

The Kalman filter provides a recursive solution for the \mmse and the final \mse. In particular, $\epsilon_{w,t}^2$, the top left element of the matrix $\Sigma_t$, denotes the \mse in estimating $w$ given the observed sequence $y^t$ and is defined as $\epsilon_{w,t}^2 = \E\sq{\cq{W-\hat{W}_t}^2}$. Therefore, $\epsilon_{w,t+1}^2$ can be recursively computed based on $\Sigma_t$ and parameters $\gamma_t$ and $c_t$. In the next section, we propose an \mdp which utilizes this recursive property in Theorem~\ref{Thm: Kalman Filter Derivation} to find the optimal parameters for the encoding scheme that minimize the final \mse, $\epsilon_{w,T}^2$.

\section{Optimal sequential linear schemes via dynamic programming}
\label{sec: dp}

In the previous section, we showed that the \mmse can be updated sequentially as a function of the encoder parameters $\{c_t,\gamma_t\}_{t=1}^T$. In this section, we provide a closed-form optimal solution of $\{c_t,\gamma_t\}_{t=1}^T$ that minimizes the \mmse at any time $t$, for a peak power constraint. We do so by formulating an \mdp, for which the parameters of the encoding scheme are modeled as an action and the corresponding \mmse is modeled as a cost (Section~\ref{sec: proposed mdp framework}), and computing the optimal action policy via dynamic programming.
(Section~\ref{sec: DP for constant peak power}). We begin with an overview of the general \mdp framework.

\subsection{General \mdp Framework}
An \mdp, in general, is described through a Markov process characterized by the tuple $\bq{s_t, a_t, \tau_t, r_t}$. An agent with the state $s_t$ takes an action $a_t$ at a cost $r_t$ and in the process transitions to the future state $s_{t+1}$ governed by the function $\tau_t$ as $s_{t+1} = \tau_t (s_t, a_t)$.
We define the optimal policy $\pi_t$ as the mapping from the state space to the action space that minimizes the expected sum of cost from any state to the final state. We define the value function $V_t\cq{s_t}$ to be the expected sum of cost if the agent follows the optimal policy $\pi_t$ from $s_t$ till the final state $s_T$. The optimal policy $\pi_t$ is computed using the Bellman equation as
\begin{align}
  \pi_t\cq{s_t} & = \underset{a_t}{\mathrm{argmin}} \left(r_t\cq{s_t, a_t} + V_{t+1}\cq{\tau_t\cq{s_t, a_t}}\right),      \nonumber \\
  V_t\cq{s_t}   & = r_t\cq{s_t, \pi_t\cq{s_t}} + V_{t+1}\cq{\tau_t\cq{s_t,\pi_t\cq{s_t}}}. \label{Eqn: Original Bellman Equation}
\end{align}

\subsection{Proposed \mdp framework}
\label{sec: proposed mdp framework}

\begin{figure}
  \begin{floatrow}
    \ffigbox{%
      % \hspace{-1em}
      \includegraphics[width=.6\textwidth]{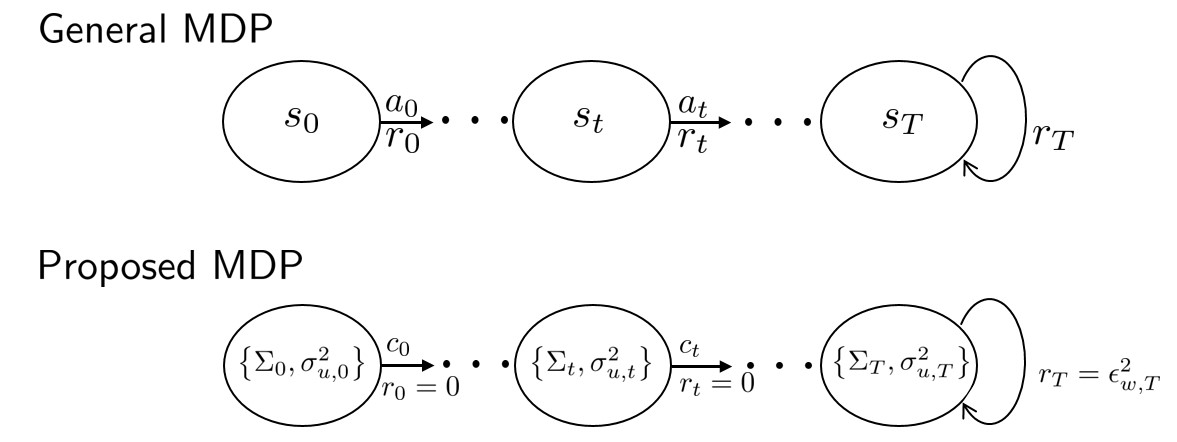}
    }{%
      \caption{\mdp Model and the analogy to the \awgn channel with feedback}
      \label{fig: MDP and Connection to Network}
    }
    \capbtabbox{%
      \begin{adjustbox}{width=.6\columnwidth,center}
        \begin{tabular}{|l|c|c|}
          \hline
          \textbf{Parameter}                   & \textbf{Symbol}           & \textbf{Value}                                    \\
          \hline
          \multirow{2}{*}{State}               & \multirow{2}{*}{$s_t$}    & \multirow{2}{*}{$\bq{\Sigma_t,\sut^2}$}           \\
                                               &                           &                                                   \\
          \hline
          \multirow{2}{*}{Action}              & \multirow{2}{*}{$a_t$}    & \multirow{2}{*}{$c_t$}                            \\
                                               &                           &                                                   \\
          \hline

          \multirow{2}{*}{Cost}                & \multirow{2}{*}{$r_t$}    & $0$ \ \ \ \ $t\neq T$                             \\
                                               &                           & $\epsilon^2_{w,T}$ \ $t=T$                        \\
          \hline
          \multirow{2}{*}{Value function}      & \multirow{2}{*}{$V_t$}    & \multirow{2}{*}{$\sum_{t=0}^T r_t$}               \\
                                               &                           &                                                   \\
          \hline
          \multirow{2}{*}{Transition Function} & \multirow{2}{*}{$\tau_t$} & \multirow{2}{*}{$g_4\cq{\cdot}$, $g_5\cq{\cdot}$} \\
                                               &                           &                                                   \\
          \hline
        \end{tabular}
      \end{adjustbox}
    }{%
      \caption{Table showing the elements of the proposed \mdp}%
      \label{Table: MDP table}
    }
  \end{floatrow}
\end{figure}

In the following, we formulate an \mdp with the encoding scheme being the policy and the \mse being the cost, which has to be minimized to get the optimal encoding scheme, as summarized in Table~\ref{Table: MDP table}. % , and described in the following.

\vspace{1em}
\noindent\textbf{State and Action\ }
We define the state $s_t$ as the tuple of the error covariance matrix and the variance of the encoder's state, i.e., $s_t = \bq{\Sigma_t, \sigma_{u,t}^2}$, and the action $a_t$ as the encoder parameter, i.e., $a_t = c_t$.
Using Theorem~\ref{Thm: Kalman Filter Derivation}, we can show the following.
\begin{theorem}
  \label{Thm: Recursion relation of St}
  \textnormal{
    Let state $s_t$ be the pair of the error covariance matrix and and the variance of the encoder's state, $s_t:= \bq{\Sigma_t, \sut^2}$, and the control action $a_t$ be the parameter $c_t$, we can define the \mdp as
    \begin{align*}
      s_{t+1} = \tau_t(s_t,a_t),
    \end{align*}
    where the transition function $\tau_t\cq{\cdot}$ is obtained from~\eqref{Eqn: ut variance final recursion in theorem} and~\eqref{Eqn: pthat and Sigmat}.
  }
\end{theorem}

\begin{proof}
  From~\eqref{Eqn: Encoder function}, we have, $u_{t+1} = u_t + c_t \tilde{y}_t$, therefore, $\sigma_{u,t+1}^2=\E\sq{U_{t+1}^2}$ can be expressed in terms of $\sigma_{u,t}^2$ and the encoder parameters as
  \begin{align}
    \sigma_{u, t+1}^2 & = \cq{1 + \gt c_t}^2 \sigma_{u,t}^2 + c_t^2\cq{\sigma_f^2 + \sigma_b^2}. \label{Eqn: ut variance final recursion in theorem}
  \end{align}
  Under the constant peak power constraint, we have $\gt=\sqrt{\frac{P}{\sut^2}}$, from which it follows that
  \begin{align}
    \sigma_{u, t+1}^2 & = \cq{1 + \sqrt{\frac{P}{\sigma_{u,t}^2}} c_t}^2 \sigma_{u,t}^2 + c_t^2\cq{\sigma_f^2 + \sigma_b^2} =: g_4(\sut^2, c_t).\label{Eqn: g4 function}
  \end{align}

  In Theorem~\ref{Thm: Kalman Filter Derivation}, we derive the recursive function $\Sigma_{t+1} = g_2\cq{\Sigma_t,\gt,\gtt,c_t}$.
  Combining $\gt=\sqrt{\frac{P}{\sut^2}}$, $\gtt=\sqrt{\frac{P}{\sutt^2}}$, and $\sutt^2 = g_4(\sut^2, c_t)$, and substituting the values for $\gt$ and $\gtt$, we can represent $\Sigma_{t+1}$ as a function of $\cq{\Sigma_t, \sut, c_t}$. We let $g_5\cq{\cdot}$ denote such a function, i.e.,
  \begin{align}
    \Sigma_{t+1} = g_5\cq{\Sigma_t, \sut^2, c_t}. \label{Eqn: g5 function}
  \end{align}

  Given that $s_t = \bq{\Sigma_{t},\sut^2}$, and $\phi_t$ includes $c_t=a_t$, we conclude $s_{t+1} = \tau_t\cq{s_t,a_t}$ from~\eqref{Eqn: g4 function} and~\eqref{Eqn: g5 function}.

\end{proof}

We now provide an intuition behind choosing $\{\Sigma_t, \sigma_{u,t}^2\}$ as the state. A naturally proposed \mdp would be to let $\Sigma_t$ denote the state and the encoding function $\phi_t$ denote the action, where $\phi_t$ is parameterized by $\bq{\gtt,c_t}$. However, $\gtt$ is not an independent variable given the peak power constraint and can be obtained from the variance of $u_t$ at the encoder, $\sigma_{u,t}^2$.

Therefore, we define our state $s_t$ to be the collection of variances $\bq{\Sigma_t, \sigma_{u,t}^2}$ and the parameter $c_t$ becomes the control action $a_t$. This allows us to define the recursion function $\tau_t$ that relates the current state $s_t=\bq{\Sigma_t, \sigma_{u,t}^2}$ and control actions $a_t=c_t$ to the future state $s_{t+1}=\bq{\Sigma_{t+1},\sigma_{u,t+1}^2}$, as we show in Theorem~\ref{Thm: Recursion relation of St}. Consequently, the problem of finding the optimal encoding scheme $\tphi_t$ is now reduced to obtaining the optimal action $\at_t$ at any given state $s_t$.

\vspace{.5em}
\noindent\textbf{Cost and Value function}\ We define the cost $r_t$ in a way that the objective function that is minimized at each step of the iteration in~\eqref{Eqn: Original Bellman Equation} is the final \mse. Specifically, we define the cost $r_t$ as
\begin{align}
  \label{Eqn: Reward definition}
  r_t =
  \begin{cases}
    0                & t \neq T \\
    \epsilon^2_{w,T} & t = T.
  \end{cases}
\end{align}
Such a definition for cost $r_t$ ensures the value function, which is the optimized sum of the cost, is always $r_T = \epsilon^2_{w, T}$. This kind of definition for cost function is common in problems where only the final cost is considered.

We define the value function as $ V_t\cq{s_t} = \epsilon_{w,T}^2\cq{s_t, \tilde{a}_{t:T-1}}$,
given any state $s_t$, $t = 0:T-1$ other than than the terminal state,
which denotes the the final \mmse if we start at state $s_t$ and undertake the optimal actions $\at_{t:T-1}$. Then the following is obtained using the Bellman equation in~\eqref{Eqn: Original Bellman Equation} by substituting the reward function as defined above. % such that
\begin{subequations}
  \begin{align}
    V_t\cq{s_t} & = \epsilon_{w,T}^2\cq{s_t, \tilde{a}_{t:T}} \label{Eqn: value function definition} \\
                & = \epsilon_{w,T}^2\cq{s_t, \tilde{a}_t, \tilde{a}_{t+1:T}}                         \\
                & = \epsilon_{w,T}^2\cq{s_{t+1}, \tilde{a}_{t+1:T}}                                  \\
                & = V_{t+1}\cq{s_{t+1}}.\label{Eqn: value function update}
  \end{align}
\end{subequations}

In summary, the key idea is to treat the $T$-step encoding process as an \mdp with the action $a_t$ representing the free parameter $c_t$ of the encoder function. The proposed \mdp and the analogy to the communication network are showcased in~\figref{fig: MDP and Connection to Network} and are summarized in table~\ref{Table: MDP table}. In the next section, we provide an algorithm for solving this \mdp.

\subsection{Solution of the \mdp via Dynamic Programming}
\label{sec: DP for constant peak power}

We presented an \mdp and defined the state $s_t:= \bq{\Sigma_t, \sut^2}$, the control action $a_t = c_t$, and the set of update equations that describe the transition $s_{t+1} = \tau_t\cq{s_t, a_t}$ in Theorem~\ref{Thm: Recursion relation of St}.
To solve the \mdp and find the optimal control action, we use the \emph{dynamic programming} which returns the analytical expressions for the $T+1$ optimal actions $\{a_t\}_{t=0}^T$ and the value function $V_t(s_t)$.

\begin{algorithm}[!htb]
  \SetAlgoLined
  \DontPrintSemicolon
  \KwOut{Optimal action $\tilde{a}_{0:T-1}$, Value function $V_{0:T}$}
  \textbf{Initialization}: At $t=T$,  $\forall$  $s_T$, $V_{T}\cq{s_T} =  \epsilon_{w,T}^2\cq{s_T,\at_T}$\;
  \For{$t = T-1, \cdots, 0$}
  {
    $\tilde{a}_t = \underset{a_t}{\mathrm{argmin}} \ V_{t+1}\cq{\tau_t\cq{s_t,a_t}}$\;
    $V_t\cq{s_t} = V_{t+1}\cq{\tau_t\cq{s_t,\at_t}}$\;
  }
  \caption{Proposed Dynamic Program}
  \label{alg: DP for peak power constraint}
\end{algorithm}

As depicted in Algorithm~\ref{alg: DP for peak power constraint}, we backward recursively evaluate the optimal actions from time $t=T$ till $t=0$, storing the value function at each instant. The algorithm initializes the value function $V_T$ at $t = T$ for all $s_T$ as $\epsilon_{w,T}^2$. The value is part of the definition of $\Sigma_T$ given in~\eqref{Eqn: Sigmat definition} within the definition of state $s_T$. The subsequent steps use the formulation proved in~\eqref{Eqn: value function update} to obtain the value function at each of the previous states $s_t$ from $t = T-1$ till $0$ while optimizing over the actions to obtain the optimal action $\at_t$.

We solve the dynamic program above and obtain the closed-form expressions for the value function $V_t(s_t)$ and the optimal action $\tilde{a_t}$ for every $t \in [0,T]$ as shown in the following.

\begin{lemma}[Solution to the \mdp]
  \label{Lemma: Solution of the dynamic program}
  \textnormal
  {
    Let $P$, $\sigma_f^2$ and $\sigma_b^2$ be the power constraint, forward and feedback channel variance respectively for the \awgn channel with feedback. Let $V_t\cq{s_t}$ be the value function for state $s_t$ at any time $t$ for the analogous \mdp, then the $V_t\cq{s_t}$ and the optimal control action $\at_t$ is given as
    \begin{align}
      V_t\cq{s_t} = \frac{-\epsilon_{uw, t}^4 + K_{T-t}\epsilon_{w, t}^2\sigma_{u, t}^2 + \epsilon_{w, t}^2\epsilon_{u, t}^2}{K_{T-t}\sigma_{u, t}^2+\epsilon_{u,t}^2},\label{Eqn: Value function computation}
    \end{align}
    while the optimal action $\tilde{a}_t$ is given as
    \begin{align}
      \tilde{a}_{t} =  -\frac{K_n\sqrt{S}\cq{\frac{\sigma_{u,t}}{\sigma_f}}}{K_n\eta_0 + \beta}, \label{Eqn: Final encoder solution}
    \end{align}
    where $n=T-t$ is the number of remaining transmissions and
    \begin{align}
      K_n & = \frac{\eta_1 K_{n-1}^2 + \eta_2K_{n-1}}{\eta_3K_{n-1}^2 + \eta_4K_{n-1}+\eta_2}
      =: f\cq{K_{n-1}}                        \label{Eqn: K_n_value}                          \\
          & = f^{n-1}\cq{K_1}, \label{Eqn: K_f_value}
    \end{align}
    with $K_1=\frac{\eta_1}{\eta_3}$, $\eta_0  = \cq{1 + S}\cq{1 + \beta}$, $\eta_1  = 1+\beta + S\beta$, $\eta_2  = \beta$, $\eta_3  = S(1+\beta)(1+S)$, $\eta_4  =  1 + \beta + S + 2S\beta$, with $S  = \frac{P}{\sigma_f^2}$ and $\beta  = \frac{\sigma_b^2}{\sigma_f^2}$.
  }
\end{lemma}

\begin{proof}
  See Appendix~\ref{App: Solution of the dynamic program} for the proof.
\end{proof}

In the next section, we obtain the sequential linear encoding scheme from the solution of the dynamic program for both noiseless and noisy feedback cases. We also derive the final \mse obtained using these schemes from the value function computed from the dynamic program.

\section{Optimal Sequential Linear schemes}
\label{sec: schemes}

In~\sect{sec: dp}, we described an \mdp analogous to a communication system with feedback and proposed a \dpp algorithm to solve it. The solution obtained was a set of $T+1$ value functions and corresponding optimal control actions. This section uses these solutions to derive closed-form expressions for the optimal encoding scheme and the \mmse and analyze its performance. In addition, we compare our proposed encoding scheme with the \sk scheme for channels with noisy output feedback and show that our scheme cannot be obtained as a trivial generalization from the \sk scheme. Furthermore, we obtain approximate expressions for \mse for large $T$ and analyze their asymptotic performances for both noiseless and noisy feedback regimes.

\subsection{Optimal sequential linear schemes}
\label{sec: optimal sequential linear schemes}

The optimal control actions obtained in Lemma~\ref{Lemma: Solution of the dynamic program} and given by~\eqref{Eqn: Final encoder solution} are used to derive the linear encoding scheme in a closed-form, as defined in~\eqref{Eqn: Encoder function}. We constructed the \mdp such that the action vector was the required coefficient $c_t$ of the encoding scheme. We use this fact to derive the encoding scheme in the following theorem.
\begin{theorem}[Closed-form solution for the optimal sequential linear encoding]
  \label{Theorem: Encoder scheme}
  \textnormal
  {
    Let $P$, $\sigma_f^2$ and $\sigma_b^2$ be the peak power constraint, forward and feedback channel variances respectively for \awgn channels with noisy output feedback. Then the optimal sequential linear encoder, denoted as $\tphi_t\cq{\cdot}$, for transmitting a message is given as
    \begin{align}
      x_{t+1} & = \gamma_{t+1} u_{t+1}, \label{Eqn: transmitted encoder scheme}
    \end{align}
    where the state of the encoder $u_{t+1}$ is updated as
    \begin{align*}
      u_{t+1} & = \cq{u_t - \frac{K_{n}\sqrt{S}\cq{\frac{\sigma_{u,t}}{\sigma_f}}}{K_{n}\eta_0 + \beta}\tilde{y} _t},
    \end{align*}
    and the scaling factor $\gamma_{t+1}$ is given as
    \begin{align*}
      \gtt = \sqrt{\frac{P}{\cq{1+\frac{K_{n}S}{K_n\eta_0 + \beta}}^2 + \frac{K_n^2S\sigma_{u,t}^2}{\cq{K_n\eta_0+\beta}^2}\cq{1+\beta}}}
    \end{align*}
    for $n = T-t-1$, and the parameters $K_n$, $\beta$, $S$ and $\eta$s are defined in Lemma~\ref{Lemma: Solution of the dynamic program}.
  }
\end{theorem}
\begin{proof}

  As shown in~\eqref{Eqn: Encoder function}, sequential linear encoders are in the form of $x_{t+1} = \gtt u_{t+1}$ where
  \begin{align*}
    u_{t+1} = u_t + c_t\yt_t \quad \text{and}\quad \gamma_{t+1} = \sqrt{\frac{P}{\sutt^2}}
  \end{align*}
  under a peak power constraint. Thus, sequential linear encoders are fully defined by the weight $c_t$ for $t \in [0,T+1]$.
  In the formulation of the \mdp in~\sect{sec: dp}, we defined the control action as $a_t = c_t$ and derived the optimal control action in~\eqref{Eqn: Final encoder solution} of Lemma~\ref{Lemma: Solution of the dynamic program}.  Substituting $c_t$ as $\tilde{a}_t$, and using~\eqref{Eqn: g4 function}, we obtain the optimal sequential linear encoding scheme.

  Therefore, the optimal encoding scheme to generate the transmitted symbol $x_t$ given feedback $\yt_t$ and the previous encoder state $u_t$ is as given as~\eqref{Eqn: transmitted encoder scheme}.

\end{proof}

\subsection{Interpretation of optimal sequential linear codes}
It is important to interpret and analyze the behavior of the optimal sequential linear scheme. To this end, we first compare the optimal sequential linear codes against the SK scheme (See~\sect{sec: sk} for a description). The following corollary shows that our scheme specializes to the \sk scheme for channels with noiseless output feedback.

\begin{corollary}[Noiseless feedback]
  \label{corollary: proving SK is same as ours}
  \textnormal
  {
    For \awgn channels with noiseless output feedback, the optimal policy $\tphi_t\cq{u_t,\yt_t}$ derived in Theorem~\ref{Theorem: Encoder scheme} coincides with the \sk scheme under a peak power constraint $P$.
  }
\end{corollary}

\begin{proof}
  This corollary immediately follows from the optimal encoding scheme obtained in Theorem~\ref{Theorem: Encoder scheme}.

  We considers a noiseless feedback channel with the forward noise variance $\sigma_f^2=1$. We have $\sigma_b^2 = 0$ i.e. $\beta:= \sigma_b^2/\sigma_f^2 = 0$.

  Using~\eqref{Eqn: transmitted encoder scheme} and substituting $\beta=0$, we get
  \begin{align}
    \tphi_t\cq{u_t,y_t} & = \sqrt{\frac{P}{\sigma_{u,t+1}^2}}\cq{u_t - \frac{\sqrt{P}\sigma_{u,t}}{1+P}y_t}          \\
                        & =\sqrt{\frac{P}{\sigma_{u,t+1}^2}}\cq{u_t - \mE[U_t|y_t]},\label{Eqn: noiseless corollary}
  \end{align}
  which coincides with the \sk scheme as in~\cite{Gallager2010}.

\end{proof}

The fact that the optimal sequential linear scheme coincides with the SK scheme is not surprising given that the SK scheme belongs to the family of sequential linear schemes. An interesting question is how/if the optimal sequential linear scheme is different from the SK scheme for channels with noisy output feedback. To answer this question, we begin by characterizing a natural extension of the SK scheme that follows the philosophy of the SK scheme.

The SK encoding scheme for noiseless feedback involves repeatedly sending the error in the receiver's estimate of the transmitted message at all subsequent steps. However, for noisy feedback channels, the encoder has access to only a {\em noisy} version of the estimate. An intuitive method would be to estimate the message at the encoder from the received noisy feedback, compute the error, and then send it as the new transmitted symbol, i.e.
\begin{align*}
  \phi_t^{SK}\cq{u_t,\yt_t} & =\sqrt{\frac{P}{\sigma_{u,t+1}^2}}\cq{u_t - \mE[U_t|\yt_t]}.
\end{align*}

The remaining question is whether the optimal sequential linear scheme coincides with the extended SK scheme. In the following, we show that the two schemes are different; the optimal sequential linear scheme strictly outperforms the extended SK scheme, bolstering the fact that our sequential linear scheme is non-trivial.

\begin{corollary}[Noisy feedback]
  \label{corollary: coefficients depend on T}
  \textnormal
  {
    For \awgn channels with noisy output feedback, where forward and feedback channel variances are given as $\sigma_f^2$ and $\sigma_b^2$ respectively, the optimal policy $\tphi_t\cq{u_t,\yt_t}$ derived in Theorem~\ref{Theorem: Encoder scheme} is not the same as the direct generalization of the \sk scheme $\phi^{SK}$, i.e.
    $\tilde{\phi}_t \neq \phi_t^{SK}$, under the peak power constraint $P$.
  }
\end{corollary}

\begin{proof}
  A strict generalization to the \sk scheme implies that the update of the encoder state $U_t$ depends on the estimate made by the encoder on the expected value for $U_t$ based on the noisy received feedback $\yt_t$ i.e.
  \begin{align}
    u_{t+1} & = u_t - \mE\sq{U_t\vert \yt_t}                                              \\
            & = u_t - \frac{\sqrt{P}\cq{\sigma_{u,t}}}{P + \sigma_f^2 + \sigma_b^2}\yt_t.
  \end{align}
  Therefore, the direct generalization of the \sk scheme is given as
  \begin{align*}
    \phi^{SK}\cq{u_t,\yt_t} = \sqrt{\frac{P}{\sigma_{u,t+1}^2}}\cq{u_t - \frac{\sqrt{P}\sigma_{u,t}}{P + \sigma_f^2 + \sigma_b^2}\yt_t}.
  \end{align*}

  To compare the extended \sk scheme, denoted by $\phi^{SK}$, and the optimal sequential linear encoding scheme $\tphi$ derived in Theorem~\ref{Theorem: Encoder scheme}, in Figure~\ref{Fig: Coefficient plot with T}, we plot the coefficient that is multiplied by feedback $\yt_0$, from the first forward transmission, for the generalized scheme $\phi^{SK}$ and the proposed scheme $\tphi_t$ as a function of the total number of transmissions $T$. % In particular, we plot the % coefficient value used to generate the transmitted symbol 
  % after receiving the first feedback. 
  The plot stays constant for scheme $\phi^{SK}$ for different values of $T$, unlike our derived scheme. This dependence on $T$ proves that our proposed scheme is not the same as the generalized \sk scheme. Also, the dependence shows that our scheme is non-trivial for noisy feedback and cannot be derived as a generalization from the \sk scheme for which the derived coefficients are constant with $T$.

  \begin{figure}[htb]
    \centering
    \begin{tikzpicture}
      \node(img){\includegraphics[scale=0.65]{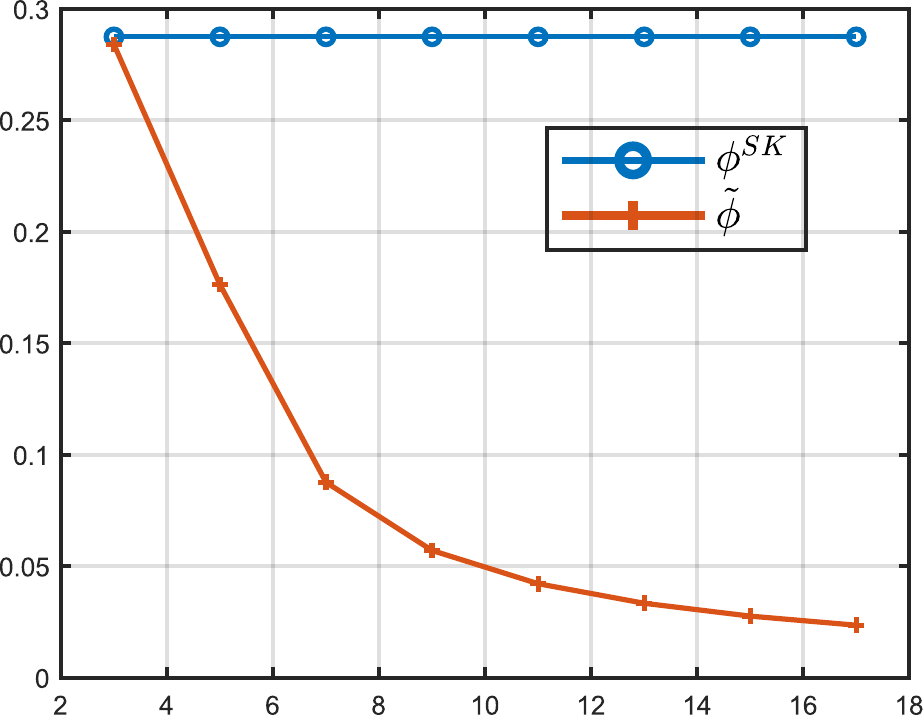}};
      \node[below=of img, node distance=0cm, yshift=1cm,font = {\small}] {Transmissions, $T$};
      \node[left=of img, node distance=0cm, rotate=90, anchor=center,yshift=-0.7cm,font = {\small}] {Coefficient Value, $c_0$};
    \end{tikzpicture}
    \caption{The first coefficient that is multiplied to $\tilde{y}_0$ (i.e., $c_0)$ over different values for the total number of transmissions $T$ with $\sigma_f = 0dB$ and $\sigma_b=-20 dB$. It can be deduced that unlike the generalized \sk scheme $\phi^{SK}$, the coefficients derived from the \dpp solution $\tilde{\phi}$ are a function of the total intended transmissions.}
    \label{Fig: Coefficient plot with T}
  \end{figure}

\end{proof}

We now turn our attention to the achievable \mse. In the following, we derive the \mse achieved by the optimal sequential linear schemes by establishing the relationship between the value functions and the final \mse.
\subsection{\mmse of the Optimal Sequential Linear schemes}
We provide a theorem to compute the closed-form expression for the \mmse at the end of $T+1$ transmissions with noisy feedback. We establish a relationship between the value functions computed in Lemma~\ref{Lemma: Solution of the dynamic program} using the \dpp algorithm in~\sect{sec: dp} with the intended final \mse.

\begin{theorem}
  \label{Thm: MSE Expression}
  \textnormal
  {
    Consider an \awgn channel with feedback with a peak power constraint of $P$, forward channel variance of $\sigma_f^2$ and feedback channel variance of $\sigma_b^2$, then \mse after $T+1$ transmissions in the estimate of the message is given by
    \begin{align}
      \text{MSE} & = \frac{\sigma_w^2}{\zeta_T}, \label{Eqn: MSE Expression}
    \end{align}
    where
    \begin{align}
      \zeta_T = \cq{1 + S + \frac{1}{K_T}}\cq{\frac{\frac{\eta_2}{K_T} + \eta_4}{\frac{\eta_2}{K_T} + \eta_1}}, \label{Eqn: Zeta Value}
    \end{align}
    with parameters $S$, $K_T$ and $\eta$'s being described in terms of the system parameters $P$, $\sigma_f^2$, $\sigma_b^2$ and $T$ as in~\eqref{Eqn: Value function computation}
  }
\end{theorem}

\begin{proof}
  A message with variance $\sigma_w^2$ can be transmitted with a peak power constraint of $P$ over a forward and feedback channel with variances $\sigma_f^2$ and $\sigma_b^2$, respectively, with the \mse after $T+1$ transmissions given by
  \begin{align*}
    \text{MSE} & = V_0\cq{\Sigma_0,\sigma_{u,0}^2}.
  \end{align*}
  This is straightforward from the assumption that all costs are $0$ except the terminal cost, i.e, \mse, and from the fact that the value function captures the sum of costs till the terminal state.

  Without loss of generality, we can assume the first transmission to be $x_0 = \sqrt{P}w$. Therefore, the receiver error variance and the encoder variance at the conclusion of the raw transmission can be expressed as
  \begin{align*}
    \Sigma_0  = \bm{\frac{\sigma_w^2}{S + 1} & \frac{\sigma_w^2}{S + 1} \\ \frac{\sigma_w^2}{S + 1} & \frac{\sigma_w^2}{S + 1}}\quad \text{and} \quad \sigma_{u,0}^2  = \sigma_w^2.
  \end{align*}
  Using the expression in~\eqref{Eqn: Value function computation} and then substituting the value function in terms of the system parameters, we obtain
  \begin{align*}
    V_0\cq{\Sigma_0,\sigma_{u,0}^2}.= \frac{\sigma_w^2}{\zeta_T}, %\label{Eqn:actual_mse_val}
  \end{align*}
  where
  \begin{align*}
    \zeta_T & = \frac{\cq{1+K_T+SK_T}\cq{K_T + \beta + \beta K_T + SK_T + 2S\beta K_T}}{\cq{K_T}\cq{K_T + \beta + \beta K_T + S\beta K_T}}
    = \cq{1 + S + \frac{1}{K_T}}\cq{\frac{\frac{\eta_2}{K_T} + \eta_4}{\frac{\eta_2}{K_T} + \eta_1}}. %\label{Eqn:zeta_actual1}
  \end{align*}
\end{proof}

\begin{figure}[!tb]
  \centering
  \begin{tikzpicture}
    \node(img){\includegraphics[width=.5\textwidth]{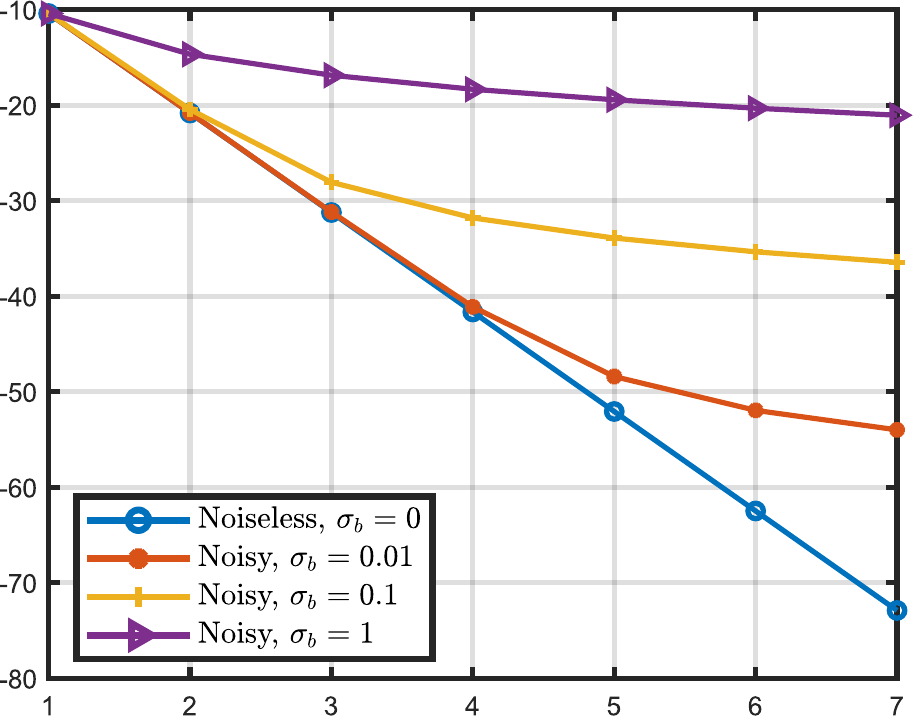}};
    \node[below=of img, node distance=0cm, yshift=1cm,font = {\small}] {Transmissions, $T$};
    \node[left=of img, node distance=0cm, rotate=90, anchor=center,yshift=-0.7cm,font = {\small}] {MSE (dB)};
  \end{tikzpicture}
  \caption{Plots showing the \mse in the estimation of the original message $w$ for $\sigma_w=1$, $P=10$, $\sigma_f^2=1$ and different values of feedback variance $\sigma_b^2$  against varying number of total transmissions $T$. The exponential nature of \mse with $T$ is evident for the noiseless case. It will be shown later that the reduction of \mse in polynomial with $T$ for noisy output feedback}
  \label{fig: MSE_DP}
\end{figure}

In~\figref{fig: MSE_DP}, we plot the \mse (dB) in the estimation of the message $w$, for $\sigma_w=1$, $P=10$, $\sigma_f^2=1$ and varying values of $\sigma_b^2$, against the number of transmissions $T$. We observe that the \mse drops exponentially for the noiseless feedback case ($\sigma_b=0$). The exponential nature is evident as the graph is linear with respect to the $T$ with the \mse in dB scale which we also show analytically (Corollary~\ref{corollary: mse decreases exponentially for noiseless feedback}).

The \mmse obtained using our proposed scheme with noisy feedback varies polynomial with $T$, but this dependence is not apparent. Therefore, we use an approximate expression to understand this dependence better in terms of $T$ for both noiseless and noisy feedback cases. The obtained approximate asymptotic expression matches the bounds that were given in~\cite{Chance2011a, Cover2003}.

\begin{corollary}[\mse for noiseless feedback]
  \label{corollary: mse decreases exponentially for noiseless feedback}
  \textnormal
  {
    Let $P$ and $\sigma_f^2$ be the peak power constraint and forward channel variance respectively for \awgn channels with noiseless feedback. The \mse in the estimate of the message of variance $\sigma_w^2$ at the receiver decays exponentially in transmissions $T$ for noiseless feedback, and the \mse is given as
    \begin{align}
      \text{\mse} = \frac{\sigma_w^2}{\cq{1+S}^T}, \text{\ where\ } S=\frac{P}{\sigma_f^2}.
    \end{align}
  }
\end{corollary}

\begin{proof}
  See Appendix~\ref{App: proof for the noiseless mse approximation}.
\end{proof}

\begin{corollary}[\mse for noisy feedback]
  \label{corollary: mse decreases polynomially for noisy feedback}
  \textnormal
  {
    Let $P$, $\sigma_f^2$ and $\sigma_b^2$ be the peak power constraint, forward and feedback channel variance respectively for \awgn channel with noisy feedback. The drop in the \mse of the optimal sequential linear schemes in estimating a message with variance $\sigma_w^2$ is approximately polynomial given as
    \begin{align}
      \text{\mse} \approx \frac{\sigma_w^2}{\zeta^*\cq{T}}, \label{Eqn: approximate mse expression}
    \end{align}
    where $\zeta^*\cq{T}$ is a polynomial function in $T$ and is given as
    \begin{align*}
      \zeta^\star = S\cq{1+\frac{1}{\beta}}T
    \end{align*}.
    with $S=\frac{P}{\sigma_f^2}$ and $\beta=\frac{\sigma_b^2}{\sigma_f^2}$.
  }
\end{corollary}
\begin{proof}
  See Appendix~\ref{App: proof for the noisy mse approximation}
\end{proof}

\begin{figure}[!tb]
  \centering
  \centering
  \begin{tikzpicture}
    \node(img){ \includegraphics[width=.5\textwidth]{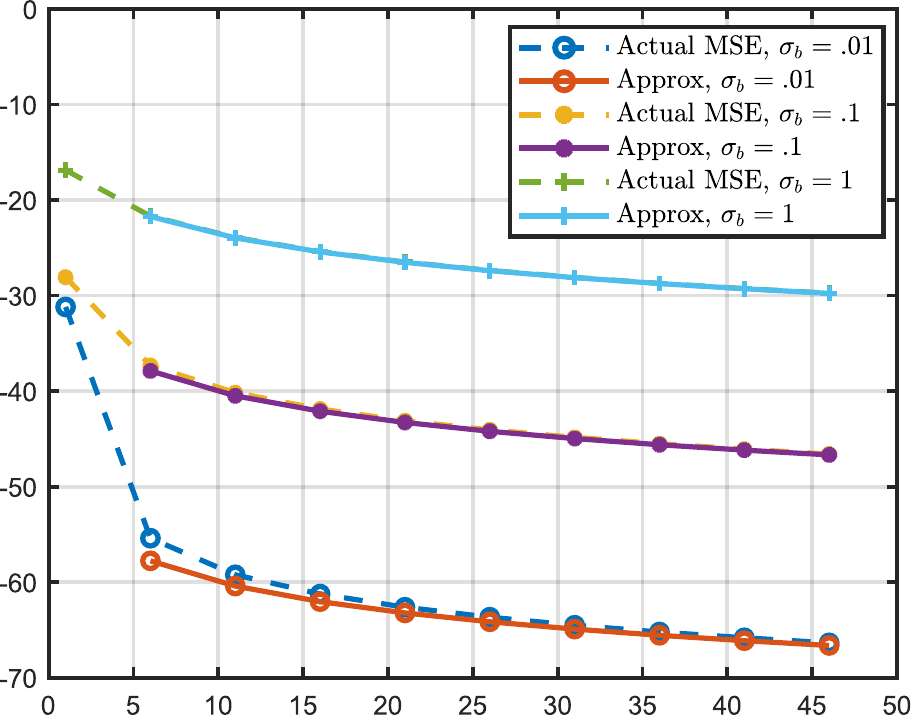}};
    \node[below=of img, node distance=0cm, yshift=1cm,font = {\small}] {Transmissions, $T$};
    \node[left=of img, node distance=0cm, rotate=90, anchor=center,yshift=-0.7cm,font = {\small}]{MSE (dB)};
  \end{tikzpicture}
  \caption{The plot shows the actual \mse and the \mse approximation for $\sigma_w=1$, $P=10$, $\sigma_f^2=1$ and different values of feedback variance $\sigma_b^2$  against varying number of total transmissions $T$. It can be seen that the polynomial approximation of \mse in~\eqref{Eqn: approximate mse expression} is tight and is better with the higher values of $T$ or with higher feedback noise variance}
  \label{fig: MSE Approximation}
\end{figure}

\figref{fig: MSE Approximation} shows the comparison of the approximated expression for \mse in~\eqref{Eqn: approximate mse expression} with the actual closed form value of the \mse as a function of $T$ obtained in~\eqref{Eqn: MSE Expression}. We empirically verify that the approximation is quite tight, and the tightness improves with the increasing value of the feedback variance $\sigma_b^2$.
\section{Total Power Constraint}
\label{sec: total power}
In the previous sections, we derived a closed-form solution for the optimal sequential linear scheme and the corresponding \mse under the {\em peak power} constraint. In this section, we derive the optimal sequential linear encoding scheme under the {\em total power} constraint. We begin with the \mdp formulated for the peak power constraint but modify the action spaces; we introduce additional parameters to represent the power allocations for each of the transmissions. We then propose a dynamic program to optimize the parameters of the encoding function and the power allocation.

\subsection{State and Action}
In~\sect{sec: dp}, we introduced an \mdp to represent our communication system with feedback as tuple $\bq{s_t, a_t, \tau_t, r_t}$ which was summarized in Table~\ref{Table: MDP table}. For the total power constraint, we introduce two new parameters in the state vector, $P_t$ and $Q_t$, where the parameter $P_t$ represents the instantaneous power used in the current transmission, while $Q_t$ is the unallocated power budget for the remaining steps, including the current, i.e.
\begin{align*}
  Q_t = \sum_{t'=t}^T P_{t'}.
\end{align*}
We also introduce a parameter $\alpha_t$ into the action vector, defined as the fraction of available power $Q_t$ that is allocated to the current transmission, i.e., $ P_t = \alpha_t Q_t$.
\begin{figure}[!htb]
  \centering
\end{figure}

\begin{figure}
  \begin{floatrow}
    \ffigbox[.6\textwidth]{%
    \includegraphics[width=.6\textwidth]{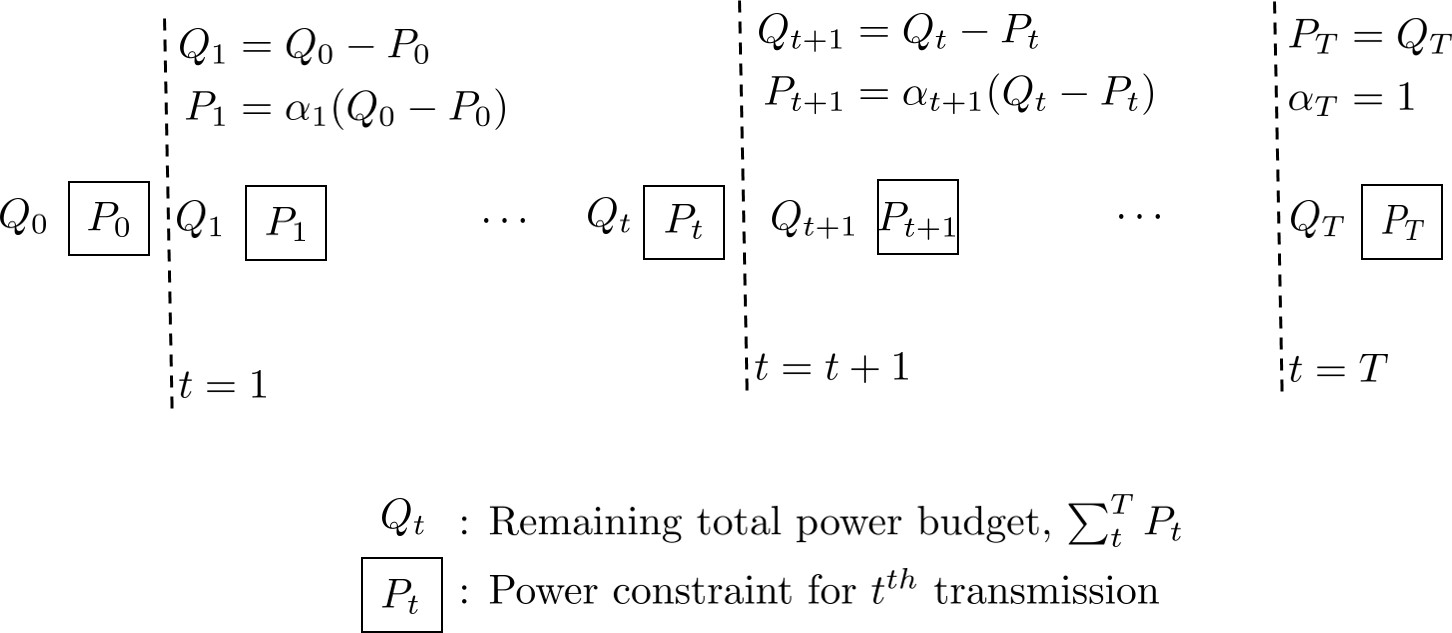}
    }{%
    \caption{Power allocation in terms of the action parameter $\alpha_{t+1}$. At time $t+1$, the power constraint is determined as $P_{t+1} = \alpha_{t+1}Q_{t+1}$ where $\alpha_{t+1}$ is optimized through the \dpp algorithm.}
    \label{fig: Total Power Illustration}
    }
    \capbtabbox[.4\textwidth]{%
      \begin{adjustbox}{width=.8\columnwidth,center}
        \begin{tabular}{|l|c|c|}
          \hline
          \textbf{Parameter}          & \textbf{Symbol}           & \textbf{Value}                                                     \\
          \hline
          \multirow{2}{*}{State}      & \multirow{2}{*}{$s_t$}    & \multirow{2}{*}{$\bq{\Sigma_t,\sut^2, P_t, Q_t}$}                  \\
                                      &                           &                                                                    \\
          \hline
          \multirow{2}{*}{Action}     & \multirow{2}{*}{$a_t$}    & \multirow{2}{*}{$c_t, \alpha_{t+1}$}                               \\
                                      &                           &                                                                    \\
          \hline
          \multirow{2}{*}{Transition} & \multirow{2}{*}{$\tau_t$} & \multirow{2}{*}{$g_6\cq{\cdot}$, $g_7\cq{\cdot}$, $g_8\cq{\cdot}$} \\
                                      &                           &                                                                    \\
          \hline
        \end{tabular}
      \end{adjustbox}
    }{%
      \caption{Table showing the elements of the proposed \mdp under the total power constraint}
      \label{Table: MDP table total power}
    }
  \end{floatrow}
\end{figure}

As illustrated in~\figref{fig: Total Power Illustration}, at any time $t$, given $P_t$ and $Q_t$, the power budget for the remaining transmissions $\sq{t+1,T}$ is $Q_{t+1} = {Q_t-P_t}$. Now, the objective is to optimize the parameter $\alpha_{t+1}$ in $P_{t+1} = \alpha_{t+1}Q_{t+1}$ to determine the appropriate power constraint $P_{t+1}$ for the ${t+1}^{th}$ transmission. This is repeated at every time instants $t\in\sq{0,T-1}$ except at time $t=T$ where $\alpha_{T}$ is unity as all the remaining power are allocated to the final transmission. Thus, the extended state and the action vectors for the \mdp are given as $s_t=\bq{\Sigma_{t}, \sut^2, P_{t}, Q_{t}}$ and $a_t=\bq{c_t,\alpha_{t+1}}$.

Let us now provide a theorem to derive the transition functions that govern the altered definitions of the state and action vectors in line with the framework provided in Theorem~\ref{Thm: Recursion relation of St}.

\begin{theorem}
  \label{Thm: Recursion relation of St with total power constraint}
  \textnormal
  {
    Let $s_t$ denote the pair of the error covariance matrix, the variance of the encoder's state and the power allocation parameters, $s_t:= \bq{\Sigma_t, \sut^2, P_t, Q_t}$, and the control action $a_t$ being the parameter $c_t,\alpha_{t+1}$. Then, $s_t$ satisfies the \mdp as
    \begin{align*}
      s_{t+1} = \tau_t(s_t,a_t),
    \end{align*}
    where the transition function $\tau_t\cq{\cdot}$ is obtained from~\eqref{Eqn: ut variance final recursion in theorem with total power},~\eqref{Eqn: Sigma transition in total power} and~\eqref{Eqn: Power transition}.
  }
\end{theorem}

\begin{proof}

  We derive recursive equations for the $\Sigma_t$, $\sut^2$ and the power parameters $P_t$ and $Q_t$. These equations eventually constitute the function $\tau_t$ that govern the transition of $s_t$.

  From~\eqref{Eqn: Encoder function}, we have $u_{t+1} = u_t + c_t \tilde{y}_t$. Therefore, $\sigma_{u,t+1}^2$ can be expressed in terms of $\sigma_{u,t}^2$, $P_t$ and the encoding function as
  \begin{align}
    \sigma_{u, t+1}^2 & = \cq{1 + \gt c_t}^2 \sigma_{u,t}^2 + c_t^2\cq{\sigma_f^2 + \sigma_b^2}, \label{Eqn: ut variance final recursion in theorem with total power}
  \end{align}
  where $\gt=\sqrt{\frac{P_t}{\sut^2}}$. From~\eqref{Eqn: ut variance final recursion in theorem with total power}, we obtain a function $g_6\cq{\cdot}$ such that $\sutt$ satisfies $\sutt^2 = g_4(\sut^2, c_t)$.

  We note that $\gtt$ is chosen to satisfy the power constraint $P_{t+1}$ at time instant $t+1$ as
  \begin{align}
    \gtt & = \sqrt{\frac{P_{t+1}}{\sigma_{u,t+1}^2}} \nonumber                                                       \\
         & = \sqrt{\frac{\alpha_{t+1}\cq{Q_t - P_t}}{\sigma_{u,t+1}^2}}. \label{Eqn: gamma relation for total power}
  \end{align}

  We already have, $\Sigma_{t+1} = g_2(\Sigma_t,\gamma_t, \gtt, c_t)$ as shown in~\eqref{Eqn: pthat and Sigmat}. From~\eqref{Eqn: ut variance final recursion in theorem with total power} and~\eqref{Eqn: gamma relation for total power}, $\Sigma_{t}$ can be expressed recursively in terms of some deterministic function $g_7\cq{\cdot}$ as
  \begin{align}
    \Sigma_{t+1} & = g_2\cq{\Sigma_t,\gt,\gtt,c_t}                             \nonumber                                     \\
                 & = g_2\cq{\Sigma_t,\sqrt{\frac{P_t}{\sut^2}},\sqrt{\frac{\alpha_{t+1\cq{Q_t-P_t}}}{\sutt^2}},c_t}\nonumber \\
                 & = g_7\cq{\Sigma_t, \sut^2,c_t,\alpha_{t+1}}. \label{Eqn: Sigma transition in total power}
  \end{align}

  In addition, from our definitions of $P_t$, $Q_t$ and $\alpha_{t+1}$, we have $Q_{t+1}=Q_t - P_t$ and $P_{t+1} = \alpha_{t+1}\cq{Q_t - P_t}$ which we can encapsulate in a function $g_8\cq{\cdot}$ as
  \begin{align}
    \bq{Q_{t+1},P_{t+1}} = g_8\cq{Q_t, P_t, \alpha_{t+1}} \label{Eqn: Power transition}
  \end{align}

  Given that $s_t = \bq{\Sigma_{t},\sut^2, P_t, Q_t}$, and $\phi_t$ includes $\cq{\alpha_{t+1}, c_t} = a_t$, we conclude $s_{t+1} = \tau_t\cq{s_t,a_t}$ from functions $g_6\cq{\cdot}$, $g_7\cq{\cdot}$ and $g_8\cq{\cdot}$.

\end{proof}

The new definition of the \mdp is summarized in Table~\ref{Table: MDP table total power}.
The cost function is defined in the same way as for the constant peak power constraint case given in~\eqref{Eqn: Reward definition}. In the later part of the section, we present this modified \dpp algorithm as a solution to the \mdp to obtain the optimal sequential linear encoder with total power constraint and then analyze the solution in noiseless and noisy feedback cases.

\subsection{Modified \dpp Algorithm}

The \dpp algorithm was presented in~\sect{sec: DP for constant peak power} to compute the optimal sequential linear encoder for the peak power constraint. The value function optimization was done over the action vector, which consisted of the coefficient $c_t$ of the received feedback. In the modified algorithm for the total power constraint, we optimize over both the coefficients $c_t$ and the fractional power allocation $\alpha_t$. We note that the action is defined as $a_t = (c_t, \alpha_t)$, and we let $a_t(1)$ and $a_t(2)$ refer to $c_t$ and $\alpha_t$, respectively.

\begin{algorithm}[!htb]
  \SetAlgoLined
  \DontPrintSemicolon
  \KwOut{Optimal action $\tilde{a}_{0:T-1}$, Value function $V_{0:T}$}
  \textbf{Initialization}: At $t=T$,  $\forall$  $s_T$, $V_{T}\cq{s_T} =  \epsilon_{w,T}^2$\;
  $\indent$\For{$t = T-1, \cdots, 0$}
  {
    $\tilde{a}_t\cq{1} = \underset{a_t\cq{1}}{\mathrm{argmin}} \ V_{t+1}\cq{\tau_t\cq{s_t,a_t\cq{1},a_t\cq{2}}}$\;
    $\tilde{a}_t\cq{2} = \underset{a_t\cq{2}}{\mathrm{argmin}} \ V_{t+1}\cq{\tau_t\cq{s_t,\at_t\cq{1},a_t\cq{2}}}$\;
    $V_t\cq{s_t} = V_{t+1}\cq{\tau_t\cq{s_t,\at_t}}$\;
  }
  \caption{Proposed Dynamic Program}
  \label{alg: DP for total power constraint}
\end{algorithm}

We solve the dynamic program for the total power constraint in Algorithm~\ref{alg: DP for total power constraint} and obtain expressions for the value functions $V_t$ and the optimal actions $\tilde{a_t}$ for every $t \in [0,T]$. The value function $V_T$ is initialized at time instant $t=T$ and the subsequent steps were followed similar to the steps in the \dpp algorithm for the peak power constraint (Algorithm~\ref{alg: DP for peak power constraint}). We derive the optimal coefficient $\cq{\at_t\cq{1}}$ and the power allocation $\cq{\at_t\cq{2}}$ through alternate optimization where they are determined at two separate steps. We use MATLAB and algebraic rearrangements to obtain solutions to the optimizations as expressions depicted in the Theorem described below.

\begin{theorem}
  \label{Thm: solution to dp for the total power case}
  \textnormal
  {
    Let $\cq{T+1}P$ be the total power available for $T+1$ transmissions, and $\sigma_f^2$ and $\sigma_b^2$ be the forward and feedback channel variance, respectively, for \awgn channels with feedback. The value function is expressed as
    \begin{align*}
      V_t\cq{s_t} = \frac{-\epsilon_{uw, t}^4 + K_{T-t}\epsilon_{w, t}^2\sigma_{u, t}^2 + \epsilon_{w, t}^2\epsilon_{u, t}^2}{K_{T-t}\sigma_{u, t}^2+\epsilon_{u,t}^2},
    \end{align*}
    which is similar to the value function that was obtained for the peak power constraint.}

  \textnormal{
    The optimal coefficient $c_t$ is given as
    \begin{align*}
      c_{t} =  -\frac{K_n\sqrt{S_i}\cq{\frac{\sigma_{u,t}}{\sigma_f}}}{K_n\eta_0 + \beta},
    \end{align*}
    where $n=T-t$ is the number of remaining transmissions and $K_n  = f_{n-1}\cq{f_{n-2}\cq{\cdots f_1\cq{K_1}}}$
    with $K_1=\frac{1+\beta + S_T\beta}{S_T\cq{1+\beta}\cq{1+S_T}}$, $\eta_0  = \cq{1 + S_i}\cq{1 + \beta}$, $S_i = \frac{P_i}{\sigma_f^2}$ and $f_i$'s are separate functions determined from the computed power constraints $P_i$'s. (Note that structure of the optimal coefficient is similar as in the peak power constraint (Lemma~\ref{Lemma: Solution of the dynamic program})).}

  \textnormal{The optimal power allocation $\alpha_{t+1}$ is obtained in a closed form for the noiseless case as
    \begin{align}
      \alpha_{t+1} = \frac{1}{T-t}, \label{Eqn: alpha power}
    \end{align}
    and for the noisy case, the solutions of $\alpha_{t+1}$ are not tractable for a general $t$. Therefore, we rely on numerical solutions for the computation of the optimal $\alpha_{t+1}$.
  }
\end{theorem}

\begin{proof}
  The proof follows from Appendix~\ref{App: Solution of the dynamic program}. The solutions are obtained in symbolic MATLAB by solving the optimization equations individually for the coefficients and the power allocation.
\end{proof}

\subsection{Interpretation}

Let us now interpret the results we obtained above regarding the power allocation for the total power constraint. We begin with an interpretation of noiseless feedback scenario. In the following, we show that the power allocation for the noiseless case is uniform across all transmissions.

\begin{corollary}
  \label{corollary: power allocation for noiseless case}
  \textnormal
  {
    For the \awgn channel with $T$ noiseless feedback transmissions of a Gaussian message, with a total power constraint of $\cq{T+1}P$, the individual power constraint is given as $P$, i.e., $\cq{T+1}P$ power uniformly spread across all $T+1$ transmissions.
  }

\end{corollary}
\begin{proof}
  We determine the individual power allocations for the noiseless case using the optimal $\alpha$ from Theorem~\ref{Thm: solution to dp for the total power case} as $\alpha_t = \frac{1}{T-t+1}$.

  From~\eqref{Eqn: alpha power}, we obtain the fractional power allocation for the first raw transmission $\alpha_0 =\frac{1}{T+1}$ by substituting $t = 0$. The total budgeted power at the beginning of the transmissions is given as $Q_0 = \cq{T+1}P$. Therefore the power allocation $P_0$ is given as
  \begin{align*}
    P_0      & = \alpha_0 Q_0 \\
    \alpha_0 & =\frac{1}{T+1} \\
    P_0      & = P.
  \end{align*}
  From \eqref{Eqn: Power transition}, the remaining power for the next transmission is given as
  \begin{align*}
    Q_1      & = Q_0 - P_0 = TP. \\
    \alpha_1 & = \frac{1}{T}     \\
    P_1      & = P.
  \end{align*}
  These steps can be repeated to obtain $Q_i$ and $P_i$, $i=2\ldots T$ and thereby the power allocations for all transmissions. These power allocations are found to be uniform as $P$.
\end{proof}

The Corollary~\ref{corollary: power allocation for noiseless case} holds as we assumed that the message $w$ is Gaussian. We note that the uniform power allocations for the noiseless case are not optimal when non-Gaussian messages like \pam symbols are considered. The power allocation with $M-$PAM messages for any number of feedback transmissions, $T$ is obtained in~\cite{Gallager2010}. It provides a non-uniform power allocation between the first raw transmission of the \pam message and the rest of the transmissions.

For noisy feedback, we use numerical techniques to solve for the power allocation for each of the transmission. This is obtained by solving the optimization $\tilde{a}_t\cq{2} = \underset{a_t\cq{2}}{\mathrm{argmin}} \ V_{t+1}\cq{\tau_t\cq{s_t,\at_t\cq{1},a_t\cq{2}}}$ in Algorithm~\ref{alg: DP for total power constraint} for the values of $\alpha_t$ which is found to be different across all transmissions. It was observed that the power allocations obtained were non-uniform across the transmissions.

\subsection{Comparison to the state-of-the-art linear schemes by Chance and Love}

The authors in~\cite{Chance2011a} have provided a linear scheme to transmit messages over an \awgn channel with noisy output feedback with a constraint on the total power which to our best of knowledge is the state of the art in the class of linear codes for such channels. The details of the scheme are discussed in Section~\ref{sec: cl}.
%The most significant contribution of our work is to derive 
In the following, we show that the analytical scheme with closed-form solutions that we derived in this paper outperforms or %for the coefficients 
% that outperform the or 
match the performance of the Chance and Love scheme.

\begin{figure}[!tb]
  \centering
  \begin{tikzpicture}
    \node(img){ \includegraphics[width=.5\textwidth]{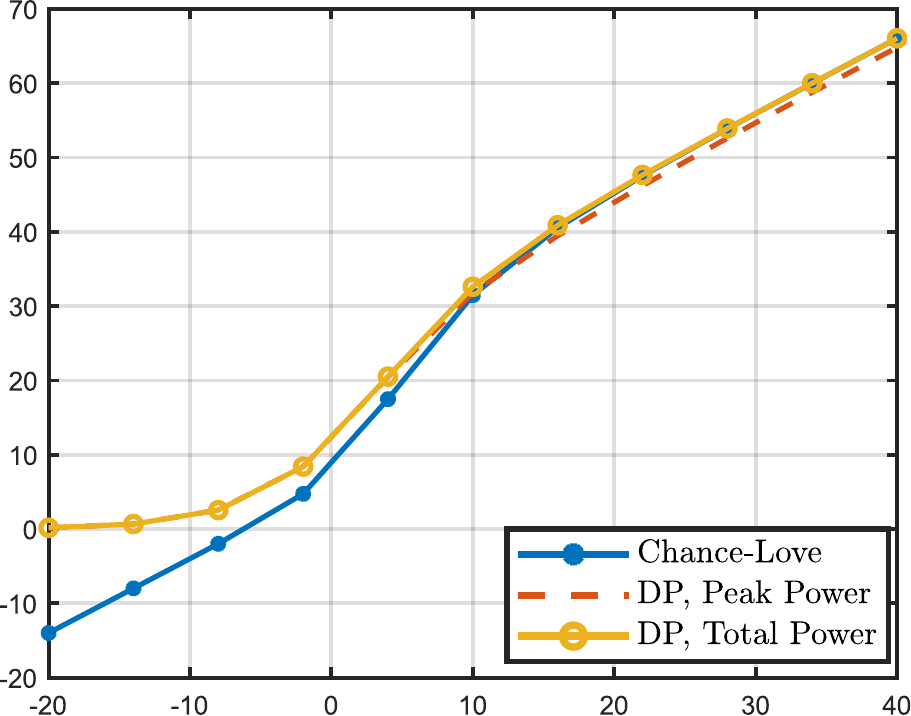}};
    \node[below=of img, node distance=0cm, yshift=1cm,font = {\small}] {Power Constraint, $P$ (dB)};
    \node[left=of img, node distance=0cm, rotate=90, anchor=center,yshift=-1cm,font = {\small}]{Received SNR (dB)};
  \end{tikzpicture}

  \caption{Comparison of the received \snr for channels with feedback between our DP scheme under peak/total power constraints and CL scheme under total power constraint~\cite{Chance2011a}. A forward noise variance of $\sigma_f^2=1$ and a feedback noise variance of $\sigma_b^2=0.01$ were assumed. At low \snr, we outperform the CL scheme in both the peak power as well as the total power constraint. At high \snr, the peak power constraint results are slightly inferior but the total power constraint results match the CL scheme. }

  \label{fig: MSE comparison with Chance Love with total power and peak power}
\end{figure}

In ~\figref{fig: MSE comparison with Chance Love with total power and peak power}, we show the comparison of the received \snr between our \dpp scheme with both peak power constraint and the total power constraint, and with the implementation of Chance and Love scheme under the total power constraint~\cite{Chance2011a}. We assume an \awgn channel with a forward noise variance of $\sigma_f^2=1$ and a feedback noise variance $\sigma_b^2=.01$. The plots were generated by varying the power constraint $P$ over $T=3$ feedback transmissions such that the peak power constraint for each transmission was $P$ while the total power constraint was $\cq{T+1}P$.

Under the total power constraint, our scheme (labeled as DP, Total Power) outperforms the CL scheme (also with the total power assumption) in the low \snr regime and has the same performance as the CL scheme at high \snr.
% We also show 
The received \snr of our scheme under the peak power constraint (labeled as DP, Peak Power) is also shown for comparison. We can see that the performance of our scheme under the peak power constraint does not degrade much with respect to the one under the total power constraint.

In summary, we show that the scheme that we propose can be obtained in closed-form for the peak power constraint which closely matches (slightly inferior) to the CL scheme while outperforming it in the low \snr regime. Our performance for the total power constraint clearly outperforms the CL scheme in the low \snr and matches the performance at high \snr.

\section{PAM Message Transmission}
\label{sec: PAM}

In the previous sections, we considered Gaussian messages to transmit over the \awgn channel with feedback. The principal motivating factor was the simplicity in determining the {\em optimal} encoding scheme within the class of sequential linear schemes with the use of the Kalman filter at the receiver as the decoder.

In this section, we consider a more practical scenario, where the transmitter has message {\em bits} (or an $M$-PAM symbol) to communicate. We empirically demonstrate that the sequential linear coding scheme that we obtained via \dpp is not optimal within the class of sequential linear schemes for such scenarios. The analysis and derivation of the optimal scheme are not straightforward when we consider messages which are not Gaussian in nature. Therefore, we use a learning-based approach; we optimize the weights of the linear sequential scheme using the backpropagation instead of \dpp. We show that the sequential linear scheme \emph{learned} for the transmission of a message bit outperforms sequential linear codes optimized for Gaussian messages, albeit by a small margin.

Concretely, we consider a setup where $w$ is the intended $m-$\pam message to be transmitted across an \awgn channel with noisy output feedback. The linear sequential encoding scheme can be represented using~\eqref{Eqn: Ut main update equation} as
\begin{subequations}
  \begin{align}
    u_0     & = w,\label{Eqn: pam encoder equation 1}              \\
    u_{t+1} & = u_t + c_t\yt_t,\label{Eqn: pam encoder equation 2}
  \end{align}
\end{subequations}
where $c_t$'s are the parameters that need to obtained. The final transmitted symbol, in line with our scheme in~\sect{sec: linear schemes}, is a scaled version of $u_t$ as
\begin{subequations}
  \begin{align}
    x_{t+1} & = \gtt u_{t+1},\label{Eqn: pam encoder equation 3}
  \end{align}
\end{subequations}
where the parameters $\gtt$ are normalized such that the total power across $\cq{T+1}$ transmissions is $\cq{T+1}P$.
In a similar manner, we represent the operation at the decoder to obtain the decoded message as a linear combination of the received symbols as
\begin{align}
  \hat{w}_T = \sum_{t=0}^Te_ty_t, \label{Eqn: pam decoder equation 1}
\end{align}
where $e_t$'s are the unknown parameters to be obtained.

The collection of parameters from the encoder and the decoder are jointly optimized through a learning-based approach using backpropagation. A batch of $m-$\pam messages ($w$'s) are generated and are passed through the system of equations in~\eqref{Eqn: pam encoder equation 1}-\eqref{Eqn: pam encoder equation 3} and~\eqref{Eqn: pam decoder equation 1} to generate the corresponding decoded messages $\hat{w}$'s in terms of the parameters. A gradient descent algorithm is used to optimize these parameters such that the \mse $\E\sq{\cq{W-\hat{W}}^2}$ is minimized. The update of the weight parameters is repeated for multiple epochs till the \mse converges to the minimal value.

\begin{figure}[!tb]
  \centering
  \begin{tikzpicture}
    \node(img){\includegraphics[width=.5\textwidth]{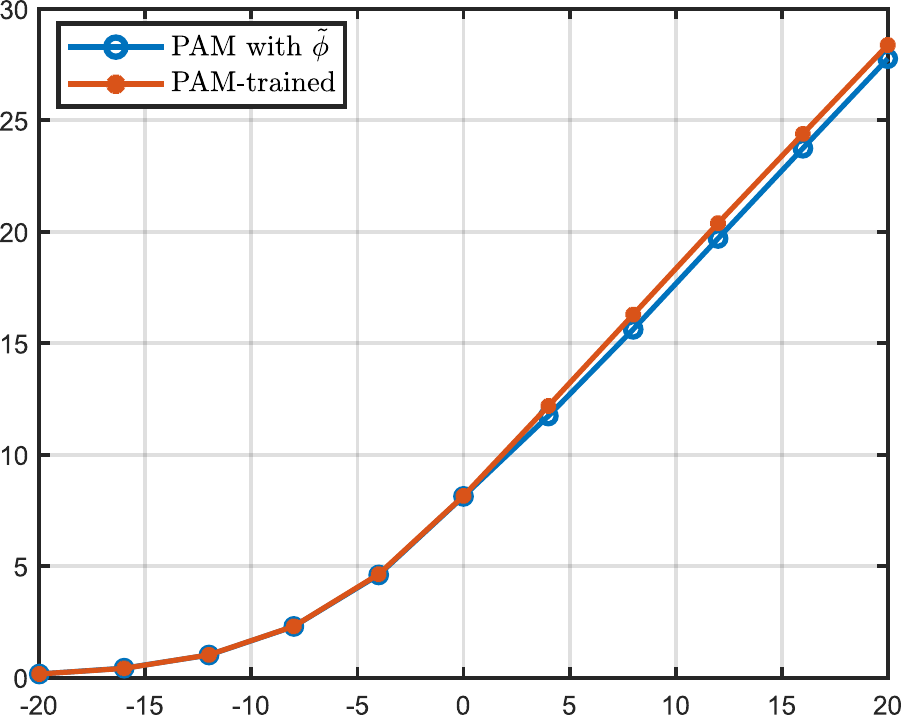}};
    \node[below=of img, node distance=0cm, yshift=1cm,font = {\small}] {Power Constraint, $P$ (dB)};
    \node[left=of img, node distance=0cm, rotate=90, anchor=center,yshift=-0.7cm,font = {\small}] {Received SNR (dB)};
  \end{tikzpicture}
  \caption{The plots compare the received \snr with $2-$\pam symbols as messages against varying power constraint $P$ in dB for an \awgn channel with feedback with forward and noise variance, $\sigma_f^2=1$ and $\sigma_b^2=1$ respectively. The improvement in the received \snr is evident when a learning based approach is used for \pam messages}
  \label{fig: MSE_PAM_DP}
\end{figure}

In~\figref{fig: MSE_PAM_DP}, we compare the performance of the sequential linear encoding scheme obtained using the proposed \dpp algorithm and the learning-based approach introduced in this section. We consider a set of $2-$\pam messages sent over the channel with $T=2$ noisy feedback transmissions. In the learning-based approach, we trained a batch of $2000$ randomly generated $2-$\pam messages and optimized using an Adam optimizer with a learning rate of $1e-3$. The performance for the scheme obtained using the \dpp approach was evaluated by generating a set of random $2-$\pam messages and then using the closed-form coefficients derived in Theorem~\ref{Theorem: Encoder scheme} to obtain the final \mse. We compute the received \snr as the reciprocal of the final \mse at the end of three transmissions. The plot shows the received \snr at the decoder for different total power constraints with the forward and the backward variances as $\sigma_f^2=1$ and $\sigma_b^2=1$. The results show a performance improvement obtained in the received \snr at high \snr under a total power constraint.

\section{Conclusion}
\label{sec: Conslusion}

We provided a novel approach using dynamic programming to design optimal sequential linear schemes for communicating a Gaussian message over \awgn channels with noisy (and noiseless) output feedback. The sequential linear scheme we introduced in this paper is a class of linear coding schemes 
for which the encoder maintains a state, which is updated based on the output feedback, and generates the transmission symbol based on the state. We showed that existing linear schemes for \awgn channels with output feedback, namely, the \sk scheme and the CL scheme, all belong to the class of sequential linear schemes. 

We then derived a closed-form expression for the optimal sequential linear scheme by formulating a novel \mdp and solving it using \dpp. 
We showed that our derived optimal sequential linear scheme outperforms the state-of-the-art CL scheme~\cite{Chance2011a} under some channel conditions for channels with {\em noisy} output feedback while matching the \sk scheme for channels with {\em noiseless} output feedback. % (\sk scheme was already known to be optimal for channels with noiseless output feedback.) 

We provided several interpretation results for the optimal sequential linear scheme and its estimation error. We showed that the derived scheme is {\em not} a straightforward generalization of the SK scheme. We represented the estimation error as a function of a number of transmissions for channels with various levels of noisy feedback. 

We also considered communicating message {\em bits} instead of a Gaussian message, for which we cannot analytically find the optimal sequential linear scheme. We presented a {\em learning-based} approach to optimize the coefficients of the sequential linear scheme directly for message {\em bits} and showed that by doing so, we can outperform the scheme with coefficients analytically optimized for a Gaussian message.

Finally, extending our approach to multi-user scenarios and using a learning-based method to directly optimize the bit error rate instead of the mean squared error would be interesting. We leave them as future work.

\bibliographystyle{IEEEtran}

%\bibliography{References}

\section{Appendix}

\subsection{Proof of Remark~\ref{Remark: SK scheme as sequentially linear}}
\label{App: SK scheme as sequentially linear}

\begin{proof}
  We show that the \sk scheme belongs to the family of sequential linear schemes.

  The sequence of receiver's \mmse estimates $\{\hat{w}_0,\hat{w}_1,\hat{w}_2,\cdots\}$ satisfies the following recursive equation
  \begin{align*}
    \hat{w}_t = \hat{w}_{t-1} + \E[U_t|y_t],
  \end{align*}
  where $u_t = w - \hat{w}_{t-1}$ denotes the error in the most recent estimate of $w$ and the estimation error $u_t$ and $y_t$ is the received signal observed as feedback at the encoder. This is derived from the fact that the error in estimate $u_t$ is orthogonal to all the observations till time $t$, $y^{0:t-1}$. Therefore, the encoding process is sequential, $u_{t+1} = u_{t} - \E[U_{t}|y_{t}]$. The transmitted symbol $x_{t+1}$ is a scaled version of $u_t$.
  \begin{align*}
    u_{t+1} & = w - \hat{w}_{t}                                           \\
            & = \cq{w - \hat{w}_{t-1}} - \cq{\hat{w}_{t} - \hat{w}_{t-1}} \\
            & = u_{t} - \E\sq{U_{t}|y_{t}},
  \end{align*}
  The transmitter sends $x_t = \gamma_t u_t$, where $\gamma_t$ denotes the power normalization constant. Under the average peak power assumption, $\gamma_t = \sqrt{P/\E[U_t^2]}$.
  The estimation error can be also represented in a recursive equation; after $t$ transmissions, the error in the estimate of $w$ reduces is $\E[(W-\hat{W})^2] = \sigma_w^2/{\cq{1+P}^T}$, which decays exponentially in $P$ (See~\cite{Gallager2010} for a detailed derivation). This scheme is analytically shown to be {\em optimal} for communication of Gaussian messages~\cite{butman1967optimum}.
\end{proof}

\subsection{Proof of Remark~\ref{Remark: chance and love as sequentially linear}}
\label{App: chance and love as sequentially linear}

\begin{proof}
  The authors in~\cite{Chance2011a} propose a scheme for channels with noisy feedback as
  \begin{align}
    x_{t+1} = \bld{F}_t\bld{\zt} +\bld{g} w, \label{Eqn: chance and love transmission equation}
  \end{align}
  where $x_{t+1}$ is the transmitted symbol, $\bld{\zt}_t$ is the noise vector, $\bld{n}_t +\bld{\nt}_t$, and $w$ is the intended message. The scheme is parameterized through the encoding matrix $\bld{F}$ which is a lower triangular Toeplitz matrix given as
  \begin{align*}
    \bm{0                                          &         & \cdots                            &                                               & 0 \\
    -\frac{1-\beta_0^2}{\cq{1-\sigma_b^2}\beta_0}  & 0       &                                   &                                               & . \\
    -\frac{1-\beta_0^2}{1+\sigma_b^2}              & \ddots  & \ddots                            & .                                             & . \\
    \vdots                                         &         &                                   &                                               &   \\
    -\frac{1-\beta_0^2}{1+\sigma_b^2}\beta_0^{T-3} & .\cdots & -\frac{1-\beta_0^2}{1+\sigma_b^2} & -\frac{1-\beta_0^2}{\cq{1-\sigma_b^2}\beta_0} & 0 \\
    }
  \end{align*}
  while $\bld{g}$ is given as
  \begin{align*}
    \bld = \sqrt{\frac{1-\beta_0^2}{1-\beta_0^{2\cq{T+1}}}}\bm{1 & \beta_0 & \beta_0^2 & \cdots & \beta_0^{T}}^T
  \end{align*}
  The value of $\beta_0$ is obtained as a solution to (31) of the paper. Here, we show that the scheme is sequential and linear in terms of the feedback and the past symbols.

  From~\eqref{Eqn: chance and love transmission equation}, we can represent the transmitted symbol at any instant $t$ as
  \begin{align*}
    x_{t+1} & = \sqrt{\frac{1-\beta_0^2}{1-\beta_0^{2\cq{T+1}}}} \beta_0^{t+1} w + \bm{-\frac{1-\beta_0^2}{1+\sigma_b^2}\beta_0^{t}            & -\frac{1-\beta_0^2}{1+\sigma_b^2}\beta_0^{t-1} & \cdots -\frac{1-\beta_0^2}{\cq{1-\sigma_b^2}\beta_0}} \bld{\zt}^{t}                                                     \\
            & = \sqrt{\frac{1-\beta_0^2}{1-\beta_0^{2\cq{T+1}}}} \beta_0^{t+1} w + \bm{-\frac{1-\beta_0^2}{1+\sigma_b^2}\beta_0^{t}            & -\frac{1-\beta_0^2}{1+\sigma_b^2}\beta_0^{t-1} & \cdots -\frac{1-\beta_0^2}{\cq{1-\sigma_b^2}}} \bld{\zt}^{t-1} +  -\frac{1-\beta_0^2}{\cq{1-\sigma_b^2}\beta_0}\zt_{t}  \\
            & = \beta_0\cq{\sqrt{\frac{1-\beta_0^2}{1-\beta_0^{2\cq{T+1}}}} \beta_0^{t} w + \bm{-\frac{1-\beta_0^2}{1+\sigma_b^2}\beta_0^{t-1} & -\frac{1-\beta_0^2}{1+\sigma_b^2}\beta_0^{t-2} & \cdots -\frac{1-\beta_0^2}{\cq{1-\sigma_b^2}}} \bld{\zt}^{t-1}} +  -\frac{1-\beta_0^2}{\cq{1-\sigma_b^2}\beta_0}\zt_{t} \\
            & = \beta_0 x_t +  -\frac{1-\beta_0^2}{\cq{1-\sigma_b^2}\beta_0}\cq{\yt_t - x_t}                                                                                                                                                                                                                              \\
            & = \cq{\beta_0 + -\frac{1-\beta_0^2}{\cq{1-\sigma_b^2}\beta_0}}x_t +  -\frac{1-\beta_0^2}{\cq{1-\sigma_b^2}\beta_0}\yt_t                                                                                                                                                                                     \\
  \end{align*}
  We represent the current transmitted symbol $x_{t+1}$ in terms of the past symbol $x_t$ and the received feedback $\yt_t$. The analysis establishes the sequential nature of the scheme.

\end{proof}

\subsection{Proof of Lemma~\ref{Lemma: Solution of the dynamic program}}
\label{App: Solution of the dynamic program}

\begin{proof}

  At time $t=T-1$, from the optimization step in Algorithm~\ref{alg: DP for peak power constraint}, $\tilde{a}_t = \underset{a_t}{\mathrm{argmin}} \ V_{t+1}\cq{\tau_t\cq{s_t,a_t}}$, we find $\at_{T-1}$ through $\frac{\partial V_T}{\partial a_{T-1}}=0$ which leads to
  \begin{align*}
    \tilde{a}_{T-1} = - \frac{\sqrt{S}\cq{\frac{\sigma_{u,T-1}}{\sigma_f}}}{\eta_0},
  \end{align*}
  where all the symbols have been defined in Lemma~\ref{Lemma: Solution of the dynamic program}. Therefore, the optimal transmission scheme is given as,
  \begin{align*}
    \tilde{\phi}_{T-1}\cq{u_{T-1},\yt_{T-1}} = \sqrt{\frac{P}{\sigma_{u,T}^2}}\cq{u_{T-1} - \frac{\sqrt{S}\cq{\frac{\sigma_{u,T-1}}{\sigma_f}}}{\eta_0}\yt_{T-1}}.
  \end{align*}
  The value function that captures the \mse if we use the optimal policy from $t=T-1$ till $T$ which is given as
  \begin{align*}
    V_{T-1}\cq{s_{T-1}} = \frac{-\cq{\epsilon_{uw, {T-1}}}^4 + K_1\cq{\epsilon_{w, T-1}\sigma_{u, {T-1}}}^2 + \cq{\epsilon_{w, T-1}\epsilon_{u, T-1}}^2}{K_1\sigma_{u, T-1}^2+\cq{\epsilon_{u,T-1}}^2}
  \end{align*}
  where
  \begin{align*}
    K_1  = \frac{1+\beta +S\beta}{\cq{1+\beta}S\cq{1+S}} = \frac{\eta_1}{\eta_3}
  \end{align*}
  We repeat the same procedure for $t = T-2$, $t = T-3$ and for any general $t$ and obtain corresponding $\tilde{\phi}$ and $V$ as
  \begin{align*}
    \tilde{\phi}_{T-2} & = \frac{\sqrt{P}}{\sigma_{u,T-1}}\cq{u_{T-2} - \frac{K_{2}\sqrt{S}\cq{\frac{\sigma_{u,T-2}}{\sigma_f}}}{K_{2}\eta_0 + \beta}\yt_{T-2}}                                          \\
    V_{T-2}            & = \frac{-\cq{\epsilon_{uw, T-2}}^4 + K_{2}\cq{\epsilon_{w, T-2}\sigma_{u, T-2}}^2 + \cq{\epsilon_{w, T-2}\epsilon_{u, T-2}}^2}{K_{2}\sigma_{u, T-2}^2+\cq{\epsilon_{u,T-2}}^2}, \\
    \tilde{\phi}_{T-3} & = \frac{\sqrt{P}}{\sigma_{u,T-2}}\cq{u_{T-3} - \frac{K_{3}\sqrt{S}\cq{\frac{\sigma_{u,T-3}}{\sigma_f}}}{K_{3}\eta_0 + \beta}\yt_{T-3}}                                          \\
    V_{T-3}            & = \frac{-\cq{\epsilon_{uw, T-3}}^4 + K_{3}\cq{\epsilon_{w, T-3}\sigma_{u, T-3}}^2 + \cq{\epsilon_{w, T-3}\epsilon_{u, T-3}}^2}{K_{3}\sigma_{u, T-3}^2+\cq{\epsilon_{u,T-3}}^2}, \\
    \tilde{\phi}_t     & = \frac{\sqrt{P}}{\sigma_{u,t+1}}\cq{u_{t} - \frac{K_{T-t}\sqrt{S}\cq{\frac{\sigma_{u,t}}{\sigma_f}}}{K_{T-t}\eta_0 + \beta}\yt_{t}}                                            \\
    V_{t}              & = \frac{-\cq{\epsilon_{uw, t}}^4 + K_{T-t}\cq{\epsilon_{w, t}\sigma_{u, t}}^2 + \cq{\epsilon_{w, t}\epsilon_{u, t}}^2}{K_{T-t}\sigma_{u, t}^2+\cq{\epsilon_{u,t}}^2},
  \end{align*}
  where
  \begin{align*}
    K_{2} & = \frac{\eta_1 K_{1}^2 + \eta_2K_{1}}{\eta_3K_{1}^2 + \eta_4K_{1}+\eta_2} = f\cq{K_{1}}.                           \\
    K_{3} & = \frac{\eta_1 K_{2}^2 + \eta_2K_{2}}{\eta_3K_{2}^2 + \eta_4K_{2}+\eta_2} = f\cq{K_{2}}.                           \\
    K_{t} & = \frac{\eta_1 K_{t-1}^2 + \eta_2K_{t-1}}{\eta_3K_{t-1}^2 + \eta_4K_{t-1}+\eta_2} = f\cq{K_{t-1}}=f^{t-1}\cq{K_1}.
  \end{align*}

  \textbf{Significance of $\mathbf{K_n}$:}
  We observe that the structure of the value expression $V_t$ with respect to the state $s_t$ remain similar barring the value of $K$ which is a function of system parameters $P$, $\sigma_f^2$ and $\sigma_b^2$ and $T$. Therefore, the value function $V_t$ and the optimal policy at any instant $t$ can be obtained by computing $K_1$ and then applying the function $f\cq{\cdot}$, $T-t-1$ times to obtain $K_{T-t}$. The expressions of $K_1$, $K_n$, and $f\cq{\cdot}$ can be obtained from Lemma~\ref{Lemma: Solution of the dynamic program}. This formulation helps us determine the value function (=\mse) and the optimal policy for any number of iterations and any value of feedback variance without worrying about the increase in complexity.

\end{proof}

\subsection{Proof to Corollary~\ref{corollary: mse decreases exponentially for noiseless feedback}}
\label{App: proof for the noiseless mse approximation}
\begin{proof}
  We provided a solution to the dynamic program in~\sect{sec: DP for constant peak power} and obtained the value function expression in Lemma~\ref{Lemma: Solution of the dynamic program}. In Theorem~\ref{Thm: MSE Expression}, we obtained the exact closed-form expressions for the \mse. The parameter $K_n$ in the expression is defined through the system parameters, the power constraint $P$, the forward noise variance $\sigma^2_f$, and the feedback noise variance $\sigma_b^2$. In order to study the variation of \mse with respect to $T$, we study the progression of $K_1$ through $K_T$.

  We begin by studying the series $K_n$, $n = 1\ldots t$ generated from the recursion function $f$. We establish that $K_n$ is a geometric series for the noiseless case where $\sigma_b=0$ and for large $n$, where
  \begin{align}
    K_n = \frac{\eta_1 K_{n-1}^2 + \eta_2K_{n-1}}{\eta_3K_{n-1}^2 + \eta_4K_{n-1}+\eta_2}.
  \end{align}
  as defined in~\eqref{Eqn: K_n_value}. We get
  \begin{align*}
    K_n                          & = \frac{1}{1+S}K_{n-1}             \\
    \text{or}\quad \frac{1}{K_T} & = \cq{1+S}^T\frac{1}{K_1}          \\
                                 & = \cq{1+S}^T\frac{\eta_3}{\eta_4}.
  \end{align*}
  By substituting in~\eqref{Eqn: MSE Expression}, we get,
  \begin{align*}
    \text{\mse}=\frac{\sigma_w^2}{\cq{1+S}^T}.
  \end{align*}

\end{proof}

\subsection{Proof to Corollary~\ref{corollary: mse decreases polynomially for noisy feedback}}
\label{App: proof for the noisy mse approximation}
\begin{proof}
  We extend the results of the noiseless case to study the variation of \mse with transmissions $T$ with noisy feedback case. We establish that $K_n$ is a harmonic series for the noisy feedback case.
  \begin{align}
    K_n = \frac{\eta_1 K_{n-1}^2 + \eta_2K_{n-1}}{\eta_3K_{n-1}^2 + \eta_4K_{n-1}+\eta_2}.\label{Eqn: polynomial appendix equation 1}
  \end{align}
  as defined in~\eqref{Eqn: K_f_value}. Now, we show that the common difference between the reciprocal terms of $K_n$ is a constant.
  From~\eqref{Eqn: polynomial appendix equation 1}, we have,
  \begin{align*}
    \frac{1}{K_n} - \frac{1}{K_{n-1}} = \frac{\eta_3K_{n-1} + \eta_4-\eta_1}{\eta_1K_{n-1} + \eta_2}.
  \end{align*}
  We observe that $K_n$ is a monotonically decreasing function. Assuming that $K_n$ becomes much smaller with increasing $n$ we get,
  \begin{align}
    \frac{1}{K_n} - \frac{1}{K_{n-1}} = \frac{\eta_4-\eta_1}{\eta_2}\nonumber \\
    \text{i.e.}\quad \frac{1}{K_T} = \frac{1}{K_1} + \frac{\eta_4-\eta_1}{\eta_2}T. \label{Eqn: KT approx harmonic expression}
  \end{align}

  This approximation holds for any value of $T$. In fact, with a high value of $T$, the difference between the exact and the approximate expression goes down to zero.

  The exact \mse expression is given in~\eqref{Eqn: MSE Expression} which can be approximated by using the harmonic progression series for $K_n$ from~\eqref{Eqn: KT approx harmonic expression}. Thus we can obtain a very close approximation for the error variance after $T$ transmissions without the need to solve the \dpp or use the function $f$ recursively, given by
  \begin{align*}
    \text{\mse} = \frac{\sigma_w^2}{\zeta_T},
  \end{align*}

  where
  \begin{align*}
    \zeta_T & = \frac{\cq{\cq{\eta_1\eta_4 - \eta_1^2}T + \eta_2\eta_3 + \eta_1\eta_4}{\cq{\cq{\eta_1\eta_4 - \eta_1^2}T + \eta_2\eta_3 + \eta_1\eta_2\cq{S+1}}}}{\eta_1\eta_2\cq{\cq{\eta_1\eta_4 - \eta_1^2}T + \eta_2\eta_3 + \eta_1^2}}                                 \\
            & = \frac{\cq{\frac{\eta_4 - \eta_1}{\eta_2}T + \frac{\eta_2\eta_3 + \eta_1\eta_4}{\eta_1\eta_2}}\cq{\frac{\eta_4 - \eta_1}{\eta_2}T + \frac{\eta_2\eta_3}{\eta_1\eta_2}+S+1}}{\frac{\eta_4 - \eta_1}{\eta_2}T + \frac{\eta_2\eta_3 + \eta_1^2}{\eta_1\eta_2}}.
  \end{align*}
  Upon further simplifications, the approximate value for \mmse for high value of $T$ can be expressed as
  \begin{align*}
    \text{\mse} \approx \frac{\sigma_w^2}{\zeta^\star}, \text{\ where\ }
  \end{align*}
  \begin{align*}
    \zeta^\star & = \cq{\frac{\eta_4 - \eta_1}{\eta_2}}T                      \\
                & = \cq{\frac{S +S\beta}{\beta}}T = S\cq{1+\frac{1}{\beta}}T.
  \end{align*}

  The bounds obtained match the results in~\cite{Cover2003} and in~\cite{Chance2011a}.

  It is worth noting that the resulting \mmse expression is described only through the system parameters. The definition for the $\eta$ parameters is provided in Theorem~\ref{Thm: MSE Expression}. This expression helps us (a) establish the bounds for our expression as was provided by other authors in their work like Weissman et al. in~\cite{Cover2003} and Chance and Love in~\cite{Chance2011a}, (b) visualize the progression of the \mmse with the increase in the parameter $T$ which was not apparent from the exact expression in~\eqref{Eqn: MSE Expression}.
\end{proof}

\end{document}